# Statistical Reconstruction of Microstructures Using Entropic Descriptors

**Ryszard Piasecki[1*] · Wiesław Olchawa[1] · Daniel Frączek[2] · Ryszard Wiśniowski[1]**

**Abstract:** We report a multiscale approach of broad applicability to stochastic reconstruction of multiphase materials, including porous ones. The approach devised uses an optimization method, such as the simulated annealing (SA) and the so-called entropic descriptors (EDs). For a binary pattern, they quantify spatial inhomogeneity or statistical complexity at discrete length-scales. The EDs extract dissimilar structural information to that given by two-point correlation functions (CFs). Within the SA, we use an appropriate cost function consisting of EDs or comprised of EDs and CFs. It was found that the stochastic reconstruction is computationally efficient when we begin with a preliminary synthetic configuration having in part desirable features.

Another option is low-cost approximate reconstructing of the entire multiphase medium beyond the SA technique. The information included in the target ED-curve was utilized for this purpose. For a given volume fraction the low-cost trial microstructures can be generated in two ways. In the first one, applied to ceramics and carbonate samples, the interpenetrating spheres generate a number of trial configurations. In the second one, with phase-EDs, here used to the sandstone sample, the overlapping superspheres do it. Both methods use a radius determined from the EDs-linked two-exponent power-law. However, the supersphere deformation parameter allows controlling of the spatial inhomogeneity of prototypical microstructures. At last, even for a hypothetical ED-curve (under reasonable assumptions), the specific microstructure can be found, if it is realizable for a given volume fraction. In general, the EDs-based methods offer a compromise between computational efficiency and the accuracy of reconstructions.

---

\* Corresponding author: Ryszard Piasecki
  piaser@uni.opole.pl
[1] Institute of Physics, University of Opole, Oleska 48, 45-052 Opole, Poland
[2] Department of Materials Physics, Opole University of Technology, Katowicka 48, 45-061 Opole, Poland

## 1 Introduction

Development of versatile statistical descriptors for extracting and analysing quantitative information from a heterogeneous material is an important branch of computational materials science. In connection to possible structure–property relationships, a few useful descriptors have been discussed (see, for example, Torquato 2002; Sahimi 2003, 2011). In particular, advanced modelling of the effective properties of a multiphase medium needs more detailed information than the knowledge of its phase-properties. The microstructural spatial characteristics of real materials should be also taken into account. For example, Liasneuski et al. (2014) studied computing the effective diffusivity of random packings of mono-sized hard spheres. The diffusivity, and similar transport properties are influenced by the heterogeneity of the packings, which is characterized and quantified by the three-point microstructural parameter $\zeta_2$ (Torquato 2002; Sahimi 2003). This parameter appears in rigorous 3th-order lower and upper improved bounds on effective properties. Its definition involves standard two- and three-point correlation functions, $S_2(r_{12})$ and $S_3$ ($r_{12}$, $r_{13}$, $r_{23}$), respectively. More specifically, Liasneuski et al. (2014) reported an approximate analytical formula for the effective diffusivity that involves the $\zeta_2$-parameter.

Generally, the disordered materials can be divided into two wide classes: spatially stationary and nonstationary media. We would like to pay attention to the former class, for which "the probability distribution function of any property does not change when shifted in space" (Tahmasebi and Sahimi 2015). In such a class, a representative sub-domain of a given two- or three-dimensional digitized sample can be chosen as the representative one for stochastic reconstruction purposes. Then, the model microstructures obtained by a low-cost stochastic reconstruction, facilitate significantly searching for the possible structure–property relationships. However, the problem of efficient reconstructing statistically equivalent three-dimensional (3D) microstructures is still a challenging task (Xu et al. 2014), especially for porous materials (see, for example, Bodla et al. 2014).

In this article, we focus on the *entropy method* of Piasecki (2011). The method was devised to the stochastic reconstructing of digitized microstructures using the hard-wall conditions (HWC). Within the SA technique, it was successfully tested on binary and greyscale images. The primary approach employs the basic overall entropic descriptor (ED) denoted as $S_\Delta$. It describes quantitatively the average degree of *spatial inhomogeneity* (Piasecki 2000a; 2000b). Its $q$-extension in Tsallis' spirit (Tsallis 1988; 2016) can be found in (Piasecki et al. 2002). It should be noticed that the usage of the hybrid pair $\{S_\Delta, C_S\}$ of EDs, where the $C_S$ measures spatial statistical complexity of Piasecki and Plastino (2010), improves the reconstructed binary microstructures. In turn, the greyscale reconstructions can be obtained via grey-level counterparts, $G_\Delta$ for grey-level *compositional inhomogeneity*, and $C_G$ for *compositional statistical complexity* (Piasecki 2009a; 2009b). In general, the cost function for the so-called unbiased hybrid reconstruction employs four hybrid-EDs, i.e., $S_\Delta$, $C_S$, $G_\Delta$ and $C_G$ (Piasecki 2011). The definition of the overall EDs is given in the next section.

The stochastic reconstructing via the SA technique can be a highly non-trivial task even for a 2D image. For example, for materials showing characteristic structural features at *different* length-scales like the two-phase laser-speckle pattern, the use of standard two-point correlation function $S_2$ alone is insufficient (Jiao et al. 2008). The authors apply the periodic boundary conditions (PBC). Even the usage of the pair $\{S_2, C_2\}$ by Jiao et al. (2009), where $C_2$ denotes cluster correlation function, cannot capture all characteristic morphological features for a given concrete sample cross-section with the irregularly shaped stone phase (Garboczi and Bentz 1998). The quantitative comparison can be found in Tahmasebi and Sahimi (2013), where a new method of reconstructing based on a cross-correlation function and a one-dimensional raster path is applied using PBC.

It is worth noticing that the limited structural information provided by the pair $\{S_\Delta, C_S\}$ of EDs and the pair $\{S_2, C_2\}$ of two-point CFs is comparatively different; see Fig. 5 of Piasecki (2009b). This observation motivated us to introduce a weighted doubly-hybrid (WDH) method (Olchawa and Piasecki 2015). The approach uses the weighted linear combination of the pairs $\{S_\Delta, C_S\}$ and $\{S_2, C_2\}$; see Sect. 3.1. When applied to a given 3D-digitized sample, the required amount of structural information can be obtained in a standard way, i.e., using a representative cross-section. However, any of EDs and CFs provides, by definition, only plane-restricted structural information. Keeping this in mind, the suitable entropic cost function, averaged per a plane and over the number of the length-scales must be used. Its definition and further details of the recent application to porous sandstone, ceramics and carbonate samples can be found in Frączek et al. arXiv:1508.03857v2.

Another option is the low-cost approximate reconstructing without use of the SA algorithm. For given phase volume fractions, one can use the information contributed by a target-curve as for instance ED-function. The origin of that curve can be diverse. For example, when we are interested in creating a synthetic microstructure, a hypothetical function of the discrete length-scales under some physical restrictions can serve as a target-curve. Typically, the needed target-curve is computed for a digitized tomography image taken for a representative real sample. In our case, a digitized surrogate microstructure was analysed.

Within this approach, a number of prototypical microstructures is generated by a model of overlapping *spheres* of a fixed radius. The radius is determined via the recently discovered two-exponent power-law (TEPL); see Olchawa et al. (2016). The comparison of the target-curve with a trial-one that corresponds to a prototypical reconstruction enables the evaluation of the cost function. The lower value of the cost function is the better statistical similarity between the target microstructure and its approximate quick reconstruction. The method was tested by Olchawa et al. (2016) on surrogate microstructures for ceramics and carbonate samples, which were earlier and carefully reconstructed within the SA (Frączek et al. arXiv:1508.03857v2). In each of the cases, among no more than fifty low-cost trials, one can select a few candidates, which are "good enough"; this point will be clarified later in Sect. 5.1. When a higher accuracy is expected, one can use the optimal approximate reconstruction as the starting configuration within the SA technique.

The second choice is an extended model of randomly overlapping *superspheres*. By definition, the shape of superspheres is a function of the so-called deformation parameter $p$; see, for example, Jiao *et al.* (2010). The meaning of parameter $p$ is clarified in Sect. 5.2. With the help of the previously developed decomposable entropic measure by Frączek and Piasecki (2014), a clear dependence of the phase inhomogeneity degree on the value of the parameter $p$ is found by Frączek et al. (2017). In this way, a leading trend in changes in a phase inhomogeneity can be forecast. Thus, making use of $p$, the fast reconstruction can be tuned to increase its accuracy. Generally, we expect, the entropy methods offer a compromise between computational efficiency and the acceptable accuracy of the SA-reconstructions.

The rest of the paper is organised as follows: In Sect. 2, we recall the definition of the overall entropic descriptors serving as a measure of spatial inhomogeneity and statistical complexity. In Sect. 3, we present the averaged objective functions for a 2D complex composite and 3D porous material. In Sect. 4, the 3D reconstruction is briefly discussed for isotropic porous sandstone, ceramics and carbonate samples. The performance of the entropic method within the framework of SA is demonstrated for the sandstone sample. Sect. 5 deals with the two variants of an approximate 3D reconstructing, without applying the SA method. In Sect. 6, we make summarizing remarks in a general context.

## 2 Entropic descriptors

It is instructive to recall a few EDs that have previously been developed to analyse some spatial features of binary and grey-scale patterns. Later on, the EDs were employed to multiscale stochastic reconstruction of various microstructures. The basis of the so-called entropy method of stochastic reconstruction is the assumption that scale sensitive statistical properties of a microstructure can be described, at least in some part, by means of the chosen EDs. Furthermore, it turned out the EDs are able to detect relatively dissimilar spatial features compared with standard two-point CFs. This opens new opportunities for developing the so-called doubly-hybrid methods in reconstructing of microstructures.

At first, we would like to point out that throughout this paper hard wall conditions are applied. Secondly, any porous media are treated as two-phase materials. Thus, the basic stochastic reconstruction of a binary $d$-dimensional sample needs only the overall entropic $S_\Delta(k; d)$-descriptor. Here, conveniently the small letter $d$ for the dimension is preferred in the formulas. The averaging per cell procedure allows us to compare the descriptor values at different length-scales $k$. According to the definition (Piasecki 2000a; 2000b), for a given binary pattern of size $L \times L$ in unit pixels or for a binary cube of size $L \times L \times L$ in unit voxels the general formula can be written as

$$S_\Delta(k; d) = \frac{S_{\max}(k; d) - S(k; d)}{\lambda(k; d)}. \tag{1}$$

This ED makes use of micro-canonical current entropy $S(k; d) = k_B \ln \Omega(k; d)$ and its maximum possible value $S_{\max}(k; d) = k_B \ln \Omega_{\max}(k; d)$. Here, Boltzmann's constant can conveniently be put equal to unity. The length-scale is given by the side length of the sampling square cell of the size $k \times k$ or the cubic one $k \times k \times k$ and sliding by a lattice constant a=1. The number of allowed positions for the sliding cell equals $\lambda(k; d) = [L - k + 1]^d$. This procedure provides, at every scale $k$, a set of cell occupation numbers $\{n_i(k; d)\}$, $i = 1, 2, \ldots, \lambda(k; d)$. In fact, for every fixed scale $k$ we analyse the auxiliary square $L_a(k) \times L_a(k)$ or the cube $L_a(k) \times L_a(k) \times L_a(k)$ of the linear size $L_a(k) \equiv (L - k + 1)k$. These "maps", composed of the sampled cells placed in a non-overlapping manner, can be treated as representative ones since they clearly reproduce at every scale $k$ a general structure of an initial sample. Such an approach allows computing the actual entropy $S(k; d)$ and the reference one $S_{\max}(k; d)$ related to the configurational actual macrostate $\mathrm{AM}(k; d) \equiv \{n_i(k; d)\}$ and the most uniform reference one $\mathrm{RM}_{\max}(k; d)$ calculated below. Keeping this in mind, the basic constraint at every scale $k$ for cell occupation numbers $n_i(k; d)$ can be written as

$$\sum_{i=1}^{\lambda} n_i(k; d) = N(k; d), \tag{2}$$

where $N(k; d)$ stands for the length scale depending on the total number of black unit objects in each of the maps. To simplify the notation we put $n_i \equiv n_i(k; d)$, $N \equiv N(k; d)$, $\lambda \equiv \lambda(k; d)$, $n_0 \equiv n_0(k; d)$ and $r_0 \equiv r_0(k; d)$.

We begin with the number $\Omega(k; d)$ of realizations of $\mathrm{AM}(k; d)$, that is the product of the ways that each of the sampled $\lambda$ cells composed of $k^d$ unit cells can be occupied with the number $n_i$ of black unit objects under the above constraint (2),

$$\Omega(k;d) = \prod_{i=1}^{\lambda} \binom{k^d}{n_i} . \tag{3}$$

In turn, the maximum possible value $S_{max}(k; d)$ is accessible for the most spatially homogeneous reference macrostate $RM_{max}(k; d) \equiv \{n_i \in (n_0, n_0 + 1)\}_{max}$, with $\lambda - r_0$ and $r_0$ number of cells occupied by $n_0 \in (0, 1, ..., k^d - 1)$ and $n_0 + 1$ of black unit objects. Thus, the following simple relation holds: $N = (\lambda - r_0)n_0 + r_0(n_0 + 1) \equiv \lambda n_0 + r_0$, where $r_0 = N$ mod $\lambda$, $r_0 \in (0, 1, ..., \lambda - 1)$ and $n_0 = (N - r_0)/\lambda$. Then, the number $\Omega_{max}(k; d)$ of microstates realizing the most uniform $RM_{max}(k; d)$, properly defined at every discrete scale $k$, is given by

$$\Omega_{max}(k;d) = \binom{k^d}{n_0}^{\lambda - r_0} \binom{k^d}{n_0 + 1}^{r_0} . \tag{4}$$

The entropic descriptor $S_\Delta$ quantifies the averaged per cell pattern's spatial inhomogeneity (a measure of configurational non-uniformity) by taking into account the average departure of a system's entropy $S$ from its maximum possible value $S_{max}$. When a system's actual entropy $S \rightarrow S_{min}$, the spatial inhomogeneity becomes maximal.

When the more detailed spatial analysis or more accurate stochastic reconstruction of the given binary pattern is required, then we recommend the simplest hybrid approach using a pair $\{S_\Delta, C_S\}$ of EDs, where the $C_S$ measures the so-called statistical complexity

$$C_S(k;d) = \frac{1}{\lambda} \frac{[S_{max}(k;d) - S(k;d)][S(k;d) - S_{min}(k;d)]}{[S_{max}(k;d) - S_{min}(k;d)]} . \tag{5}$$

Now, the minimum possible value $S_{min}(k; d) = k_B \ln \Omega_{min}(k; d)$ is available for the most spatially inhomogeneous reference macrostate $RM_{min}(k; d) \equiv \{n_i \in (0, 0 < n < k^d, k^d)\}_{min}$, with $\lambda - q_0 - 1$ of empty cells, at the most one cell occupied with the number $n$ of black unit objects and $q_0$ of fully occupied cells. The obvious relation holds: $N = n + q_0 k^d$, where $n = N$ mod $k^d$, $q_0 = (N - n)/k^d$ and $q_0 \in (0, 1, ..., \lambda - 1)$. The number of proper microstates is therefore

$$\Omega_{min}(k;d) = \binom{k^d}{0}^{\lambda - q_0 - 1} \binom{k^d}{n} \binom{k^d}{k^d}^{q_0} \equiv \binom{k^d}{n} . \tag{6}$$

The entropic descriptor $C_S$ is able to quantify the statistical complexity (a measure of complex behaviour) by taking simultaneously into account the average departures of a system's entropy $S$ from both its maximum possible value $S_{max}$ and its minimum possible value $S_{min}$. When these two departures are similar to each other, the statistical complexity is maximal. Additionally, in Appendix a possible formal extension of EDs-based procedure revealing higher "order" structural information is given. A remark about the difference between the two hybrid descriptors, $S_\Delta$ and $C_S$, is also presented.

Obviously, we can use the same ideas to obtain grey-level counterparts of the EDs, which are useful for multi-phase materials. They can also be applied even to the binary pattern that is encoded in two ways: (a) the typical one (0 = the black phase, 1 = the white phase) and (b) the greyscale fashion (0 = the black phase, 255 = the white phase); compare Piasecki (2011).

## 3  The averaged objective functions

The quality of the stochastic reconstructing of porous microstructures making use of EDs, which we present in this paper, can be illustrated on 2D and 3D binarized microstructures. Since the two distinct cost functions using the SA technique are applied, we describe them separately.

### 3.1  The cost function for 2D complex composite material

A digitized cross-section with a linear size of 170 pixels of the piece of concrete (Garboczi and Bentz 1998) with nanometre-sized pores and centimetre-sized aggregates has been used for testing different reconstruction methods (Jiao et al. 2009; Tahmasebi and Sahimi 2013; Olchawa and Piasecki 2015). The binary target pattern is composed of non-uniform arrangement of irregular aggregates, which are relatively big in comparison to the size of the whole pattern. According to our experience for the entropy based reconstruction methods within the SA, the use of an objective function composed of different EDs leads to a higher structural accuracy at a wider range of length scales. Thus, the usage of four hybrid EDs for the mentioned above types of patterns was a preferred choice in Piasecki (2011). It was applied also to the stochastic reconstruction of complex labyrinth patterns by Piasecki and Olchawa (2012). The next significant improvement is the replacing one of the entropic pairs by a pair of distinct CFs and adding two respective weighting coefficients. The WDH method has been applied to islands, aggregates or compact clusters of various shapes and poly-dispersed in sizes by Olchawa and Piasecki (2015). The approach is concisely presented below.

   The modified objective function can be described as average "energy" per a descriptor. Here, the objective multi-scale function is the weighted sum of squared and normalized differences between the values of binary EDs related to the actual configuration and the target pattern, and similarly, between the values of the CFs for the black phase. The differences are normalized with respect to the maximal values of target EDs and CFs marked with the superscript '0'. To simplify further notation we will omit the dimension $d$ wherever it does not lead to misunderstanding. Correspondingly, the normalized EDs differences can be written as

$$\widetilde{S}_{\Delta}(k) - \widetilde{S}_{\Delta}^{0}(k) \equiv [S_{\Delta}(k) - S_{\Delta}^{0}(k)]/\max S_{\Delta}^{0}(k) \ , \tag{7a}$$

$$\widetilde{C}_{S}(k) - \widetilde{C}_{S}^{0}(k) \equiv [C_{S}(k) - C_{S}^{0}(k)]/\max C_{S}^{0}(k) \ . \tag{7b}$$

In a similar way, the related differences can be written for the correlation $S_2$ and the cluster $C_2$ functions.

   For the purposes of making a comparison, the energy $E$ is additionally averaged over the number of considered scales. The final formula used by Olchawa and Piasecki (2015) reads

$$E = \frac{1}{4n}\left\{ \alpha \sum_{k\,odd}^{L} \left[ \left(\widetilde{S}_{\Delta}(k) - \widetilde{S}_{\Delta}^{0}(k)\right)^2 + \left(\widetilde{C}_{S}(k) - \widetilde{C}_{S}^{0}(k)\right)^2 \right] \ + \right.$$
$$\left. (1-\alpha) \sum_{r=0}^{L/2-1} \left[ \left(\widetilde{S}_{2}(r) - \widetilde{S}_{2}^{0}(r)\right)^2 + \left(\widetilde{C}_{2}(r) - \widetilde{C}_{2}^{0}(r)\right)^2 \right] \right\} \tag{8}$$

Here, the parameter $0 < \alpha < 1$ and the two coefficients $\alpha$ and $1-\alpha$ are treated as the weighting factors. For test purposes, the values of $\alpha = 0.1, 0.2, \ldots, 0.9$ were considered in each of the twenty runs. To each of the series, a different random seed has been chosen. Notice that for the EDs and CFs, the identical number $n = L/2$ of length scales appears.

Instead of a standard random initial configuration, a synthetic one with the same number of compact clusters as that of the target is created. To carry it out, one of the two developed approaches within cellular automata frame can be chosen (Olchawa and Piasecki 2015). This is the key point for speeding-up microstructure reconstruction within the SA technique. The program procedure allows requiring the same values for the reconstructed and target interface. The process terminates when three conditions related to the accuracy, interface and number of clusters are fulfilled. The competition of the doubly-hybrid pairs ensures considering a wider spectrum of morphological features. We present part of the results with the neutral weighting factor $\alpha = 0.5$ (Olchawa and Piasecki 2015). Here, for comparison purposes, the lineal-path $L(k)$ function, additionally computed for both the concrete sample cross-section and its reconstruction, is shown in Fig. 1a. In both of the insets, the white phase corresponds to the cement paste, while the black (or grey) phase of the concentration 0.51 represents the stones. In Fig. 1b the pair of EDs is compared, the solid lines and the symbols for the target pattern and its reconstruction, respectively. Correspondingly, in Fig. 1c, a similar comparison for the pair of CFs is given. For completeness, in Figs. 1b, c, the two hybrid EDs and the two hybrid CFs for the initial synthetic pattern are marked by the dashed lines (red and blue online, respectively in each of pairs). One can observe that the accuracy of the WDH approach is positively verified. Notice that for the weighting parameter $\alpha$ close to zero there prevails the contribution from the pair of the correlation functions $\{S_2, C_2\}$. In this case the computation time becomes longer compared to the opposite situation, i.e., when $\alpha$ is close to one and the pair of entropic descriptors $\{S_\Delta, C_S\}$ comes into play. It should be mentioned that all the computations, including the lineal-path functions, were done under HWC.

## 3.2 The cost function for 3D porous material

Now, we apply the ED-based method to stochastic reconstructions of porous material under the condition that only a single two-dimensional input image of an entire three-dimensional sample can be used in order to reconstruct it. This is one of the most difficult and time-consuming computational problems in the reconstruction in particular for samples with large sizes. Therefore, to accelerate the process of reconstruction we apply merely one entropic descriptor $S_\Delta(k; d=2)$, the quantitative measure for average spatial inhomogeneity of a system composed of finite-size objects. (Of course, using the reconstruction obtained in this way as the starting configuration for the more challenging and time-consuming WDH approach, one can obtain an improved reconstruction.)

This ED can naturally be applied to evaluation of statistical similarity of any two structures, say 'A' and 'B'. The more statistically similar structures 'A' and 'B' are, the closer the values of the corresponding curves $S_\Delta(k; \text{A})$ and $S_\Delta(k; \text{B})$ become, and reversely. The statistical "distance" between such two curves can be calculated as the sum over length scales of the squared differences $[S_\Delta(k; \text{A}) - S_\Delta(k; \text{B})]^2$. It should be underlined here that using the information contained in the hypothetical target-curve $S_\Delta^T(k)$, given under reasonable assumptions, in principle the specific microstructure can be found, if it is realizable for a given volume fraction. Somewhat a similar situation appears for a generalization of the Debye 2D random-medium function (hypothetical medium with short-range correlations) discussed by Cule and Torquato (1999), *cf.* Eq. (8), and described by Eq. (12.19) in Torquato (2002).

However, the multiscale 3D reconstructing procedure is more efficient when we begin with the synthetic three-dimensional configuration. It is randomly generated with the overlapping spheres of a radius, depending on the structure under consideration (Frączek et al. arXiv:1508.03857v2 [cond-mat.stat-mech]).

The general idea of our approach is quite simple. Let us introduce a Cartesian coordination system with the origin in a corner of the cube of the linear size $L$ and the axes oriented along its edges. The sample is treated as a set composed of three subsets of each of the $L$ planes. The three subsets contain the stacks of the $L$ planes being cross-sections of the 3D sample perpendicular to the $x$, $y$ and $z$-axis, respectively. Our final 3D reconstruction is acceptable when any plane of this set is statistically similar to the 2D input image treated as the target pattern. To be precise, we define the entropic cost function per plane, i.e., the averaged objective function

$$E_{avg} = \frac{1}{3L} \sum_{p=1}^{3L} E_p \,. \tag{9}$$

Here, $E_p$ denotes the sum of squared and normalized differences between the values of normalized EDs related to a current configuration of the plane $p$ and the target pattern. The latter can also be selected from a larger parent image as a representative sub-domain (this is the case here). Then, the sum is averaged over the number $N_k$ of the length scales considered,

$$E_p = \frac{1}{N_k} \sum_{k=k_0}^{k_1} \left[ \widetilde{S}_\Delta(k, p) - \widetilde{S}_\Delta^T(k) \right]^2 \,. \tag{10}$$

The EDs are normalized with regard to the maximal value of the target entropic descriptor $S_\Delta^T(k_{max})$ marked with the superscript 'T'

$$\widetilde{S}_\Delta(k, p) - \widetilde{S}_\Delta^T(k) \equiv [S_\Delta(k, p) - S_\Delta^T(k)] \big/ S_\Delta^T(k_{max}). \tag{11}$$

The maximal value of the target entropic descriptor is reached at the scale $k_{max}$. The standard definition of $S_\Delta(k)$ given by (1) is applied with $d = 2$.

The further stages of our approach are presented in the next section. Generally, during the reconstruction process for the chosen number of loops with the assigned increasing lengths, the value of the cost function $E_{avg}$ considerably decreases. The SA scenario for the temperature loops ensures the proper limiting behaviour of minimized $E_{avg}$. At the same time, also the interface $I$ per plane, denoted here as $<I>$, is actively monitored.

## 4  Three dimensional reconstruction within the SA technique

### 4.1  Selection of a representative subdomain

Each of the representative square subdomains of the linear size $L = 300$ is chosen from the corresponding larger 2D parent image (sample cross-section). These 2D parent images of isotropic porous samples (sandstone of the size $700 \times 700$, ceramics and carbonate of the sizes $500 \times 500$) were obtained from researchers at CSIRO[*]. Firstly, using the $L \times L$-sampling cell, we detect for each of the parent images on what length scale $k \equiv k_{max}$ the first peak of $S_\Delta(k)$, which is usually also a global maximum, appears the most frequently. Then, the list of locations of the corresponding sampling cells is sorted over the slightly fluctuating volume fractions of the phase under analysis. Among the cases with the volume fraction closest to the parent image, we choose one, for which also the value of the plane "interface" is proportional to that of the parent image. As a result, we obtain a representative 2D target pattern of the size $300 \times 300$ with the proper length scale $k_{max}$, the appropriate phase volume fraction and the corresponding

---

interface value. Now, the 2D target-curve $S_\Delta^T(k)$, as a function of the length scale $k$ can be calculated. In this paper, we concentrate on the case of sandstone sample, mainly.

## 4.2 Generation of starting 3D configuration

It may be convenient to start with the reversed phase colours in the 2D target image. Therefore, for the present samples, the volume fraction of the black phase after the reversing of colours is always less than 0.5. Now, let us consider a cube of the size $L^3$, composed of only black phase unit voxels. To generate an initial random 3D configuration with the needed volume fraction of the black phase, instead of white single voxels we use the overlapping spheres composed of white voxels and having a fixed radius $R$. The positions of the sphere centres are drawn with a uniform probability distribution inside the cube and in the external zone of an appropriate width. The width of the zone is determined in such a way that at least one voxel of every white sphere must be an internal voxel of the cube. Close to the ending of cutting the white wholes (or pieces) from the black 3D matrix, some trials may be rejected until the same volume fraction of the black phase is obtained like in the target. This manner of porous configuration generating can be named the balls-procedure.

The entropic cost function $E_{avg}$, described by Eqs. (9-11) with the $S_\Delta(k)$ given by Eq. (1) for $d = 2$, shows a feature that is particularly useful for the stochastic reconstruction purposes. Let us generate, using the balls-procedure, that initial 3D configurations for a series of discrete values of $R$ taken from a wide enough interval. Then, for the associated family of $E_{avg}$-curves the approximate local minimum of $E_{avg}$ appears for a characteristic discrete value of the radius $R$. In this way, the optimal starting 3D configuration can be prepared immediately in a few seconds, using the detected $R$-value.

In fact, at this stage, an approximate 3D reconstruction of interest to us is obtained. For instance, the corresponding initial value $E_{avg}$(start) is less than $68.8 \times 10^{-3}$ for sandstone ($49.0 \times 10^{-3}$ for ceramics and $86.1 \times 10^{-3}$ for carbonate). Since our algorithm is the most efficient in creating aggregates, a higher value of the initial interface, compared to the target one, is preferred here. The further work is done making use of the SA method. We point out that only a limited number of scales $k$ can be taken into account without significantly worsening the reconstruction quality. Within the present approach, we use every second scale, $k = 2, 4,\ldots,$ until the half of the image size. There are two reasons for this. First, we are interested in morphological features, which are typical of smaller length scales, i.e., not greater than $L/2$. A similar range of length scales is characteristic of other methods, e.g., for two-point correlation functions (Torquato 2002). On the other hand, the computations performed for 75 scales instead of 150 is obviously much more computationally efficient and still satisfactory enough as well.

## 4.3 Simulated annealing technique

At this stage, we employ the SA approach, which should further minimize the starting entropic cost function $E_{avg}$(start). After the interchange of the voxels (here one can say – the pixels on the planes), the new trial configuration equivalently called the system's state, is accepted with the probability $p(\Delta E_{avg})$, according to the standard Metropolis acceptance rule

$$p(\Delta E_{avg}) = \min[1, \exp(-\Delta E_{avg}/T)].$$
(12)

Here, $\Delta E_{avg} = E_{avg,\,new} - E_{avg,\,old}$ is the difference in "energy" between two successive states, which is related to the changes on 6 planes each time. Upon the acceptance, the trial pattern becomes a current one, and the evolving procedure is repeated. A fictitious temperature $T$ follows the cooling schedule, $T(l)/T(0) = \gamma$, with the chosen parameter $\gamma = 0.80$, the initial temperature $T(0) = 10^{-8}$, the $l$-th temperature loop of increasing length and fixed number of the loops $l = 26$.

However, some reconstruction details are non-standard. Having determined the "worst" $\omega$-plane with the maximal energy $E_P$ among $3L$ planes, we are in position to start the preferential selection of two voxels of different phases, called here "biased mode". If the volume fraction of the black phase on the $\omega$-plane is higher (lower) than of the target one, then a voxel drawn should be of black (white) colour before the exchanging. To accelerate computation, the voxel of the remaining colour is drawn in such a way that it does not belong to the three planes connected with the first voxel. In addition, for symmetry reasons, at least for one of the three planes associated with the second voxel, the volume fraction should change toward the target value, too.

Let us denote the numbers of black n.n. and black n.n.n. for the white centre as $w_{nn}$ and $w_{nnn}$, and similarly, for the black centre as $b_{nn}$ and $b_{nnn}$. By treating n.n. on an equal footing with n.n.n., one can ensure their equal contributions. Thus, the appropriate weights are introduced in the "neighbouring" rules for every two pixels of different phases randomly selected:

$$(10b_{nn} + 3b_{nnn} < 10w_{nn} + 3w_{nnn}) \quad \text{and} \quad (b_{nn} \leq w_{nn}) \tag{13a}$$

or

$$(10b_{nn} + 3b_{nnn} = 10w_{nn} + 3w_{nnn}) \quad \text{and} \quad (b_{nn} < w_{nn}) \,. \tag{13b}$$

At this stage, our algorithm favours the lowering of the averaged interface $<I>$ by creating aggregates. When the current value of $<I>$ is below that of $I_{\text{target,}}$ then the rules (13a, b) are not active. Then, the entirely random selection of two voxels of different phases, called here the "unbiased mode", favours the raising of the $<I>$ value. Thus, we apply the following switching: when the current $<I>$ value exceeds the value of $I_{\text{target}}$, the biased mode comes into play, while in the opposite case – the unbiased one.

Generally, during the reconstruction process for the chosen number of loops with the assigned increasing lengths, the value of the cost function $E_{avg}$ considerably decreases. The SA scenario for the temperature loops ensures the proper limiting behaviour of the minimized $E_{avg}$. At the same time, also the interface $I$ per plane denoted here as the $<I>$ is actively monitored. If all temperature loops are completed, the reconstruction terminates enabling a comparison with the results obtained in similar conditions for other samples. The method is tested in the next section on three 2D single cross-sections for 3D different porous microstructures.

### 4.4  An illustrative example of stochastic reconstruction

For each of the 2D parent images of isotropic porous sandstone (ceramics and carbonate) samples, selected earlier as the target patterns, the representative subdomains of the size $L \times L$ with $L = 300$ and the porous phase fraction $\phi = 0.19731$ (0.38144 and 0.14381), respectively, were the only allowable input to reconstruct the needed 3D structures. Here, the corresponding target curves suggest that the sandstone is the most representative sample because of its adequately high spatial uniformity. On the other hand, the carbonate turned out to be the worst, as was confirmed also by a simple observation with a naked eye. Surprisingly, the obtained results indicate the ceramics to be the most difficult sample to reconstruct with our method.

However, even using the simplest version of the entropic approach within the same SA scenario, the obtained results are quite satisfactory. As it can be seen in **Table 1**, the corresponding outcomes differ in the final ratio $E_{avg}(\text{start}) / E_{avg}(\text{end})$ as well as in the numbers of accepted MC-steps.

**Table 1** Some of the results for the entropic $S_\Delta(k)$-descriptor based the multiscale statistical reconstructions of 3D porous samples from the related single cross-sections.

| Sample | $E_{avg}$(start) | $E_{avg}$(end) | $E_{avg}$(start) / $E_{avg}$(end) | # of accepted MC steps |
|---|---|---|---|---|
| Sandstone | $68.8 \times 10^{-3}$ | $0.161 \times 10^{-3}$ | 429 | $1.1 \times 10^6$ |
| Ceramics | $49.0 \times 10^{-3}$ | $1.330 \times 10^{-3}$ | 37 | $1.4 \times 10^6$ |
| Carbonate | $86.1 \times 10^{-3}$ | $0.557 \times 10^{-3}$ | 155 | $1.3 \times 10^6$ |

For the CPU Intel 7 (3.3 GHz) without code parallelization, the overall computation time for sandstone (ceramics and carbonate) samples was about 4.3 h (5.4 and 5), respectively. In Fig. 2a the solid curve (the red online) corresponds to the overall entropic descriptor $S_\Delta(k)$ for the sandstone target pattern, see the upper inset. The open circles (the blue online) refer to the plane, see the bottom inset, being one of the 900 planes with the $E_p$ energy, which is the nearest to the final $E_{avg}$ energy (after the 3D reconstruction). In turn, for illustration purposes, in Fig. 2b the 3D exterior view of the reconstructed sandstone sample is presented. The porous phase is the green online while the rest of the sandstone is transparent. The corresponding illustrative cross-sections are illustrated in Fig. 2c. Again, the porous phase is the green online. For better visibility, the rest of the sandstone is the grey online this time. Similar quality illustrative results are obtained for the ceramics and carbon samples (Frączek et al. arXiv:1508.03857v2).

One point needs a short explanation. We need to remind that our method was primarily developed to apply to materials composed of solid phases, while here it has been applied to the porous media. Nevertheless, among the final reconstructions, the fraction of isolated solid clusters in the carbonate sample was of the $10^{-4}$ order, while for the remaining two cubes the related fractions are by two orders lower. However, the program current version can easily be modified to avoid those unrealistic effects.

On the other hand, we have checked the possible impact of isolated solid clusters, making use of a simple algorithm. The main point is how to consolidate the black phase, preserving the overall isotropy of the samples? This condition can be fulfilled by selecting randomly one of six main directions in order to make a shift of the given isolated cluster. As expected, the values of $S_\Delta(k)$ calculated for each of the final 3D configurations without any isolated cluster are practically identical with the counterparts referred to all the reconstructed cubes.

In addition, the method of multiscale entropic reconstruction can be enriched by considering also other entropic descriptors. For example, for a two-phase microstructure, instead of a single overall ED one can employ two phase-EDs obtained by splitting the overall entropic measure by Frączek and Piasecki (2014). This could allow considering more details about spatial arrangement for each of the phases. Thus, further improving the accuracy of the 3D reconstructions is possible. Such an approach is applied in Sect. 5.2 although for a different method of approximate reconstruction with an additional parameter. One can expect that the use of the phase-EDs expands capabilities of the standard entropic method within the SA.

## 5. The approximate 3D reconstruction beyond the SA technique

To reconstruct an entire two-phase medium for a given phase volume fraction, two versatile approaches are described in the next sections. They are based on the knowledge of assumed (model) or computed the $S_\Delta(k)$-function. The needed structural information contains the target ED-curve itself. For example, when we are interested in a synthetic microstructure, a hypothetical function of the discrete length-scales under some physical restrictions can serve as a target-curve. In turn, for a real sample of the linear size $L$, the target-curve is computed at all scales $k = 1, \ldots, L$ for the corresponding digitized tomography image.

On this basis, one can readily generate a series of statistically similar approximate reconstructions. We note that our method can be applied to discover what kind of a "synthetic" microstructure can be matched to the proposed hypothetical target ED-curve. The approach, in a simpler version, uses only interpenetrating spheres randomly distributed. The more advanced model of randomly overlapping superspheres extends the possible variants of the prototypical microstructures. Moreover, it also allows controlling the spatial inhomogeneity of each phase. Both of them make use of the two-exponent power-law mentioned below.

## 5.1 The TEPL and the model of interpenetrating spheres

The balls-procedure to generate the initial synthetic 3D configuration that was described in brief in Section 4.2 can be improved considerably. Formerly, a specific starting configuration was randomly generated with the overlapping balls of a radius depending indirectly on the structure considered. Recently, we proposed an approximate reconstruction of random heterogeneous microstructures, using the two-exponent power-law of Olchawa et al. (2016). This rule originates from the entropic descriptor that is a multi-scale measure of spatial inhomogeneity for a given microstructure. The corresponding formula for TEPL can be written as

$$<\max S_\Delta(\phi, R; L)> = A(L)\,\phi^{0.41} R^{q(L)}, \tag{14}$$

where $\log_{10} A(L) = 21.8 / L + 0.37$, $q(L) = -45.5 / L + 2.96$ and $L$ is the linear size of a voxel-cube. The formula relates the arithmetic average of maximums of the spatial inhomogeneity denoted as $<\max S_\Delta(\phi, R; L)>$ to the variables $\phi$ and $R$. Here, $\phi$ means the volume fraction of the matrix porous-phase called porosity, $1 - \phi$ denotes the complementary fraction of the solid-phase and $R$ is the radius of interpenetrating spheres of the solid phase, which are randomly distributed on a regular lattice.

For randomly generated configurations, we expect the following behaviour: the smaller the radius $R$ is, the lower average spatial inhomogeneity should appear, so $<\max S_\Delta(\phi, R; L)>$ should be lower, too. Such a behaviour can be observed if $q(L) > 0$ and consequently, Eq. (14) can be used safely when $L > 15$. On the other hand, for larger linear sizes, i.e., for $L \to \infty$, the formula is simplified to the limiting form

$$<\max S_\Delta(\phi, R)> \cong 2.34 \; \phi^{0.41} R^{2.96}. \tag{15}$$

The key point is to obtain a number, say $N$, of low-cost but adequate trial three-dimensional configurations. To do this, we employ the aforementioned model of overlapping solid-phase spheres but at the present stage, the fixed value of the radius $R$ is unknown. However, having calculated target's entropic descriptor, we know the values of $\max S_\Delta(k; \text{target})$ and the related length scale, $k_{\max}(\text{target})$. This allows temporary substituting in Eq. (14) the obtained maximum value instead of the average value of the random variable. Now, the needed $R$-value can be specified directly from Eq. (14). This way guarantees that for the generated current $N$-trial configurations, the simulated $\max S_\Delta(\phi, R; L)$-values should be distributed around the value of $\max S_\Delta(k; \text{target})$. Thus, for different seeds, any number of low-cost model configurations for given $\phi$ and $L$ can be obtained very easily. All we have to do is to select among them a final configuration, for which the $\max S_\Delta(\phi, R; L)$-value and the maximum related length scale $k_{\max}$ are the closest to their target counterparts, i.e. $\max S_\Delta(k; \text{target})$ and $k_{\max}(\text{target})$. This approach was tested on surrogate samples of ceramics and carbonate. In each of the cases, about fifty low-cost trials revealed a few qualitatively good enough candidates to select the optimal one. By this we understand the acceptance of the trial curves with proper location of the maximum value and similar in shape to the target curve. If necessary, see the next section, also the quantitative evaluation of the statistical distance between such two curves can be obtained by minimizing a sum over length-scales of the squared proper differences.

Exemplary low-cost but approximate reconstructions for ceramics and carbonate samples with the linear size $L = 300$ were presented by Olchawa et al. (2016). At this stage, when a better accuracy is expected, one can use the final reconstructions as the starting configurations to the standard SA technique. The case of sandstone is addressed in the next section, where the interpenetrating randomly distributed superspheres will be applied.

## 5.2 The prototypical microstructures with a controllable spatial inhomogeneity

A wide variety of real stochastic composites can be studied by means of prototypes of multiphase microstructures with a controllable spatial inhomogeneity. To create them, we propose a versatile model of randomly overlapping superspheres of a given radius and deformed in their shape by the parameter $p$. In this section, the meaning of parameter $p$ is different from that used previously in Sect. 3.2 to describe the index of a plane. A $d$-dimensional super-sphere with the radius $R$ can be defined as

$$|x_1|^{2p} + |x_2|^{2p} + ... + |x_d|^{2p} \leq R^{2p}, \tag{16}$$

where $x_i$ are Cartesian coordinates, $i = 1, ..., d$, and $p \geq 0$ is the deformation parameter responsible for the particle shape deformation from that of a $d$-dimensional sphere ($p = 1$). The parameter $p$ allows changing the shape from convexity to concavity as $p$ passes downward through 0.5; (see Fig. 1, Jiao *et al.* 2010).

On the other hand, recently the overall multiphase entropic descriptor $S_\Delta$ has been decomposed into '$w$' phase entropic descriptors, $S_{i,\Delta}$ with $i = 1, 2, ..., w$, which were denoted earlier as $f_{i,\Delta}$ by Frączek and Piasecki (2014). The $i$th-phase entropic descriptor per cell for a multiphase material build of '$w$' phases is defined by the formula

$$S_\Delta(k) = \sum_{i=1,...,w} (f_{i,\max} - f_i) / \lambda = \sum_{i=1,...,w} f_{i,\Delta}(k) \equiv \sum_{i=1,...,w} S_{i,\Delta}(k), \tag{17}$$

where $f_i = k_B \ln \Omega_i \equiv S_i$ denotes the $i$th phase Boltzmann entropy and $f_{i,\max} = k_B \ln \Omega_{i,\max} \equiv S_{i,\max}$ means its maximal theoretical value. We recall only the basic details. In what follows, we set $k_B = 1$. The $\Omega_i(k)$ is the corresponding number of realizations for a 'non-equilibrium' actual macrostate (AM) defined as a set $\{m_i(\alpha, k)\}$ of occupation numbers by the $i$th phase for overlapping sampling $\lambda$-cells of the size $k \times k \times k$ in voxels, $\alpha = 1, 2, ..., \lambda(k)$. Similarly, $\Omega_{i,\max}(k)$ describes the number of realizations for the 'equilibrium' reference macrostate (RM) that relates to a maximally uniform configuration at a given discrete length-scale $k$. The sum of $S_{i,\Delta}$ over the phases equals exactly the overall $S_\Delta$.

With the help of the decomposable entropic measure, a clear dependence of the overall as well as the $i$th phase inhomogeneity degree on the values of the parameter $p$ is demonstrated (Frączek et al. 2017). For the chosen values of $p$, the $i$th-phase inhomogeneity degree evolves at different scales for two and three-phase examples of prototypical microstructures. Indeed, a main trend in changes of the phase inhomogeneity can be predicted. Therefore, the approach can be very effective in preparing improved starting configurations for reconstructing real materials.

In the former section, the simple model to generate low-cost preferred configurations making use of randomly overlapping spheres (with $p = 1$, using the present notation) of a radius specified from the TEPL was briefly described. Now, we apply the similar model but with the superspheres defined by Eq. (16). Since the superspheres with a fixed radius $R$ are free to overlap, clusters of various sizes, shapes and volumes are created. The key point is the use of the shape deformed parameter $p$. In this way, a powerful tool for creating, in a controllable way, prototypes of random multi-phase microstructures is at hand. Here, we show how the improved approach works, using an example of porous medium.

Its 3D reconstruction based on the overall $S_\Delta(k)$-descriptor was presented briefly in Sect. 3.2. It was obtained earlier using the corresponding single 2D input image of real porous sandstone (Frączek et al. arXiv:1508.03857v2). The carefully reconstructed microstructure for sandstone (as a cube of the linear size $L = 300$ in voxels) is assumed to be the *surrogate* 3D-target for our test purpose. Under this working assumption, a part of the linear size $L^* = 150$ of porosity 0.19715, which is very close to the original one 0.19731, can be conveniently separated to speed-up the present test. For this smaller cube, the target phase-EDs and the overall ED are computed. Knowing the maximum of the latter one, i.e., max $S_\Delta(k_{max}=30; L=150) = 125.561$, one can easily obtain – via the TEPL formula given by Eq. (14) – the approximated value of radius $R = 5.223$. Making use of this $R$-value, a number of trial approximate microstructures can effortlessly be obtained.

In Fig. 3a, the solid lines represent the target phase-EDs, the $S_{1,\Delta}(k)$ for the solid phase and the $S_{2,\Delta}(k)$ for the porous one. Among twenty-five low-cost trials, we selected one that was good enough for our purposes, i.e., *quantitatively* the best. The dashed lines correspond to the phase-EDs, which are attributed to this approximate best reconstruction. The bottom long-dashed curve relates to the $S_{1,\Delta}(k; reconstr.)$ for the solid phase while the upper short-dashed line refers to the $S_{2,\Delta}(k; reconstr.)$ for the porous one. As we see, around the highest peak, the both dashed lines are located under the solid curves for scales $k < 70$. This is a point when the impact of shape deformation parameter $p$ on evolution of the phase spatial inhomogeneity comes into play. In general, the greater the parameter $p$ is, the higher the phase spatial inhomogeneity appears. That is what we need. In this case, for $p = 1.25$ we obtain a shift toward the higher degree of spatial inhomogeneity. In Fig. 3b, the open squares (circles) illustrate the increase in the accuracy of the initial approximate reconstruction for the solid (porous) phase of sandstone. It should be noted that additionally the radius $R$ was slightly lowered to 5.07 in order to move on the left the maxima of both "curves" marked with the symbols. The shorter the distance between the solid bottom (upper) curve and the corresponding open squares (circles) is, the better the statistical similarity is between the target microstructure and the approximate reconstruction $p$-improved, *cf.* Fig. 3b.

In turn, for the three microstructures of the porous phase (represented online by the green colour while the solid phase is transparent) Fig. 4 illustrates the three-dimensional exterior views of the related cubes of size 150×150×150 in unit voxels and Fig. 5 shows the corresponding cross-sections, in the following main cases: (a) for the target microstructure, (b) for the selected approximate microstructure with $R = 5.223$ and $p = 1$, and (c) for the $p$-improved prototypical microstructure with slightly lowered $R = 5.07$ and raised $p = 1.25$.

For comparison of the previous cases, Fig. 6 presents the distributions computed in a chosen direction using a cubic measurement cell of the side length $L = 32$ sliding by a unit lattice constant: a) for local percolation probability as a function of local porosity and in the inset, the corresponding local porosity distributions, and b) bimodal histograms of local conductivities with distinct peaks, the first one around the mode related to percolating cells and in the inset, for the non-percolating case, where the narrowed width of bins has been used. The electrical conductivities of the solid phase and the material filling of the pore space were fixed as $10^{-6}$ and 1 (in arbitrary units), respectively. To compute the local conductivities, the real-space renormalization group approach developed by (Shah and Ottino 1986) was used. In addition, in Fig. 6c the pore size distribution functions and average pore sizes are shown. The same cases as the previous ones - the target, approximate and $p$-improved microstructure - are distinguished by different symbols: the open circles (the red online), the open squares (the grey online) and the filled circles (the blue online), respectively. One can observe some local fluctuations, which are natural for the chosen linear size $L = 32$ of the sampling cell. Regardless of that, the results are acceptable and support the last two comparatively fast approaches (although approximate) to statistical reconstructing, especially to large-size three-dimensional porous samples.

Summarizing briefly, in order to increase the reconstruction accuracy of the first method, the quantitatively selected approximate reconstruction can be tuned within the alternative approach making use of the shape deformation parameter $p$.

## 6. Summary

A collection of stochastic reconstruction methods based on various entropic descriptors was described with a focus on binary porous materials. Among them, we distinguish two general groups. In the first one, the approaches using the SA technique and hybrid descriptors provide accurate stochastic reconstructions. However, their computation time is relatively long particularly in the double-hybrid approach. The second group include the approaches without the use of time-consuming SA algorithm. Instead, they utilize models of overlapping spheres or superspheres to generate various prototypical microstructures. Nevertheless, they provide efficiently approximate stochastic reconstructions of acceptable accuracy.

The utility of our point of view was tested on samples of different materials including porous ones. The microstructural information provided by entropic descriptors is essential at all the length-scales. For this reason, the entropic descriptors are useful for different types of real materials. This is confirmed by exemplary reconstructions presented in this article.

## 7. Appendix

Omitting in formulas dimension $d$ for simplicity and using Eq. (1), the definition of the spatial statistical complexity $C_S$ given by Eq. (5) can be rewritten as

$$C_S(k) = \frac{1}{\lambda(k)} \frac{[S_{\max}(k) - S(k)][S(k) - S_{\min}(k)]}{[S_{\max}(k) - S_{\min}(k)]} \equiv S_\Delta(k)\,\gamma(k) \tag{A1}$$

where

$$\gamma(k) \equiv \frac{S(k) - S_{\min}(k)}{S_{\max}(k) - S_{\min}(k)} \quad . \tag{A2}$$

Thus, the $C_S$ can be treated as the $S_\Delta$ corrected by a factor $\gamma$ linear in $S$. However, the whole descriptor $C_S$ is a nonlinear function of $S$ in contrast to the $S_\Delta$ alone that is linear in $S$. This has some meaning for improving the quality of the stochastic reconstructions when the hybrid cost function is employed.

On the other hand, for $0 < \gamma(k) < 1$, the power expansion of the function $[1 - \gamma(k)]^{-1}$ includes $\gamma$-terms of any order. Thus, taking into account all the components in the hypothetical series the simple relation is fulfilled at every scale $k$

$$S_\Delta(k)[1 + \gamma(k) + \gamma^2(k) + \ldots] = \frac{S_\Delta(k)}{1 - \gamma(k)} \equiv \frac{S_{max}(k) - S_{min}(k)}{\lambda(k)} . \tag{A3}$$

The hybrid EDs-pair, $\{S_\Delta(k), C_S(k)\} \equiv \{S_\Delta(k), S_\Delta(k)\gamma(k)\}$, has been frequently employed by the entropic method of multiscale statistical reconstruction. One can suppose that the next term of the third order in $S$, i.e., the $S_\Delta(k)\gamma^2(k)$, may provide an additional information useful for the SA approach. At last, the slightly better structural accuracy could be potentially obtained by using the triplet of the hybrid EDs, $\{S_\Delta(k), S_\Delta(k)\gamma(k), S_\Delta(k)\gamma^2(k)\}$.

## Figure captions

**Fig. 1** The accuracy of the WDH reconstruction of a concrete microstructure supported by comparing different descriptors. **a** Comparison of the lineal-path $L(k)$ function for the target concrete cross-section (the solid line) which is depicted in the upper inset (adapted from Jiao et al. 2009) with that of the reconstructed microstructure with $\alpha = 0.5$ (the filled circles), see the bottom inset (Olchawa and Piasecki 2015). In the both insets, the white phase corresponds to the cement paste, while the black (or grey) phase of the concentration 0.51 represents the stones. **b** Comparison of the EDs. **c** Comparison of the CFs. For completeness, the dashed lines (red and blue online) describe the corresponding descriptors computed for the initial synthetic pattern.

**Fig. 2** The quality of the $S_\Delta$-based method of reconstructing 3D porous microstructure, using a single cross-section (Frączek et al. arXiv:1508.03857v2). **a** Comparison of the $S_\Delta(k)$ function for the target cross-section of sandstone, the solid line (the red online), depicted in the upper inset with that for one of the 900 planes in the reconstructed cube, the open circles (the blue online), shown in the bottom inset. The selected plane is the optimal one, since the associated $E_p$ energy is the nearest to the final $E_{avg}$. **b** The exterior view of the reconstructed cube, where the porous phase is green while the rest of the sandstone is transparent. **c** The corresponding illustrative cross-sections.

**Fig. 3** Illustration of the efficiency of low-cost approximate reconstructing via TEPL (Olchawa et al. 2016) of a three-dimensional porous microstructure of the linear size $L = 150$ in unit voxels using the target phase-EDs, the $S_{1,\Delta}(k)$ for the solid phase, the bottom solid line (the black online) and the $S_{2,\Delta}(k)$ for the porous phase, the upper solid line (the green online). **a** The long (short) dashed lines correspond to the phase-EDs attributed to approximate reconstruction for the solid (porous) phase. **b** The open squares (circles) correspond to the phase-EDs (Frączek et al. 2017) attributed to the $p$-improved reconstruction for the solid (porous) phase.

**Fig. 4** Related to **Fig. 3**, three-dimensional outside views of the corresponding microstructures of the porous phase exclusively (represented online by the green colour). **a** The target microstructure. **b** The selected approximate microstructure with $R = 5.223$ and $p = 1$. **c** The $p$-improved prototypical microstructure with the slightly lowered $R = 5.07$ and raised $p = 1.25$.

**Fig. 5** The same as in **Fig. 4** but for the corresponding cross-sections.

**Fig. 6** For comparison purposes other structural statistics are presented. The open circles (the red online) correspond to the target, the open squares (the grey online) refer to the trial microstructure and the filled circles (the blue online) relate to the $p$-improved case. **a** Local percolation probability as a function of local porosity $\phi$ for a cubic measurement cell of the side length $L = 32$. In the inset, the corresponding local porosity distributions are depicted. **b** Bimodal histograms of local conductivities with distinct peaks, the first one around the mode ascribed to percolating cells and in the inset, for the non-percolating case. The electrical conductivity of the solid phase is $10^{-6}$ and 1 (in arbitrary units) for the material filling of the pore space. **c** The pore size distribution functions versus pore radius (in voxels). Additionally, the average pore sizes are given.

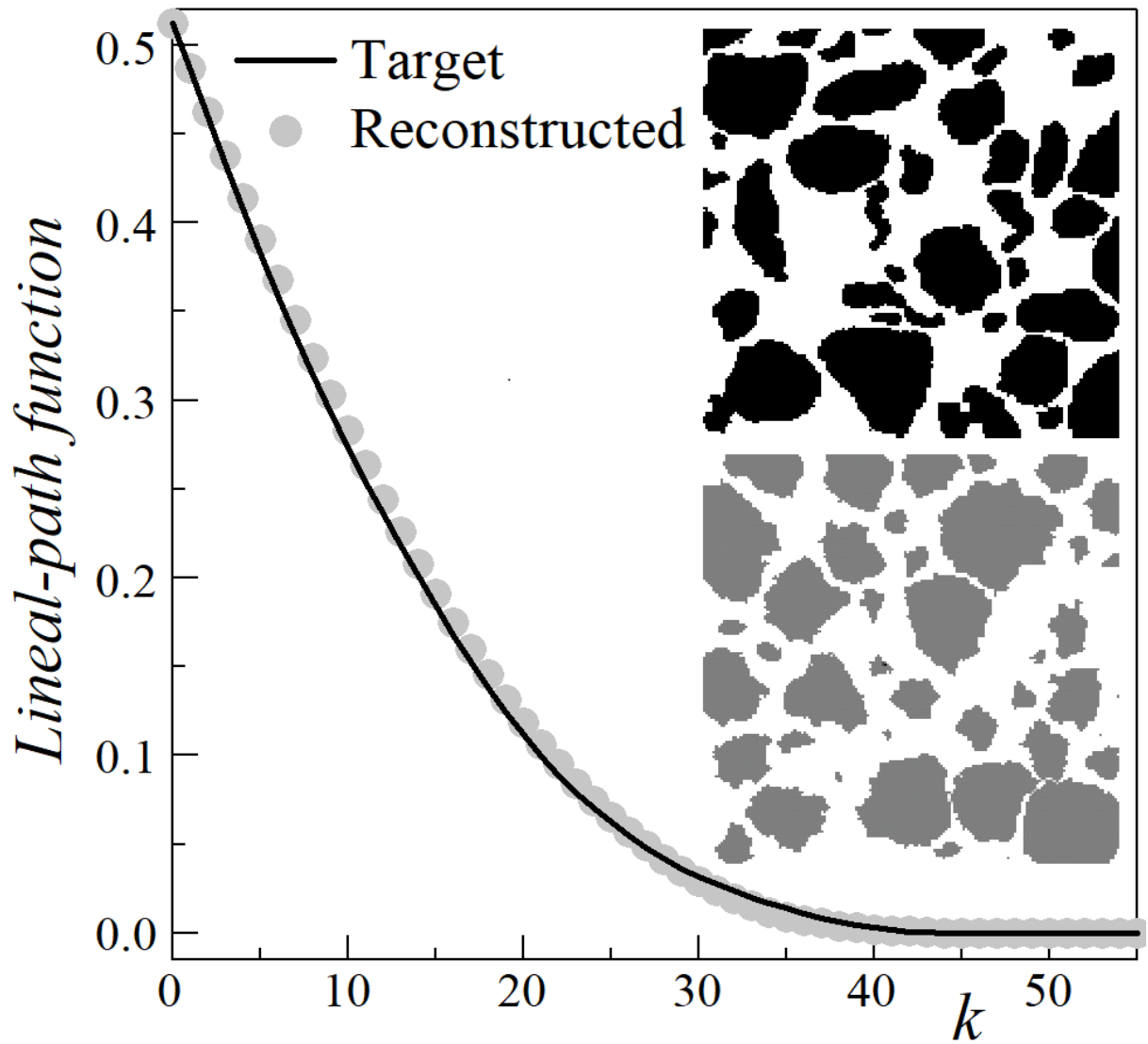

**Fig. 1a**

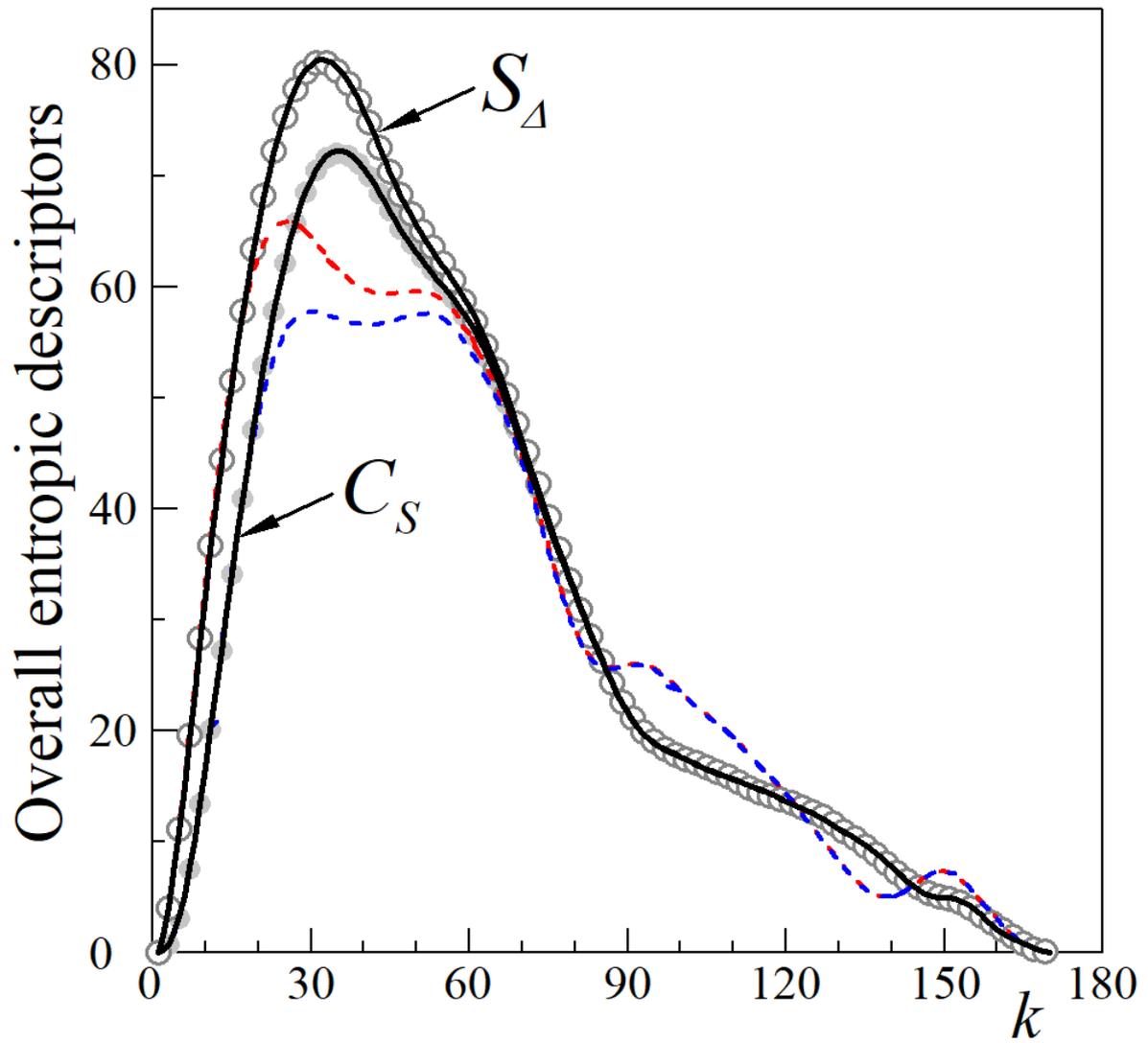

**Fig. 1b**

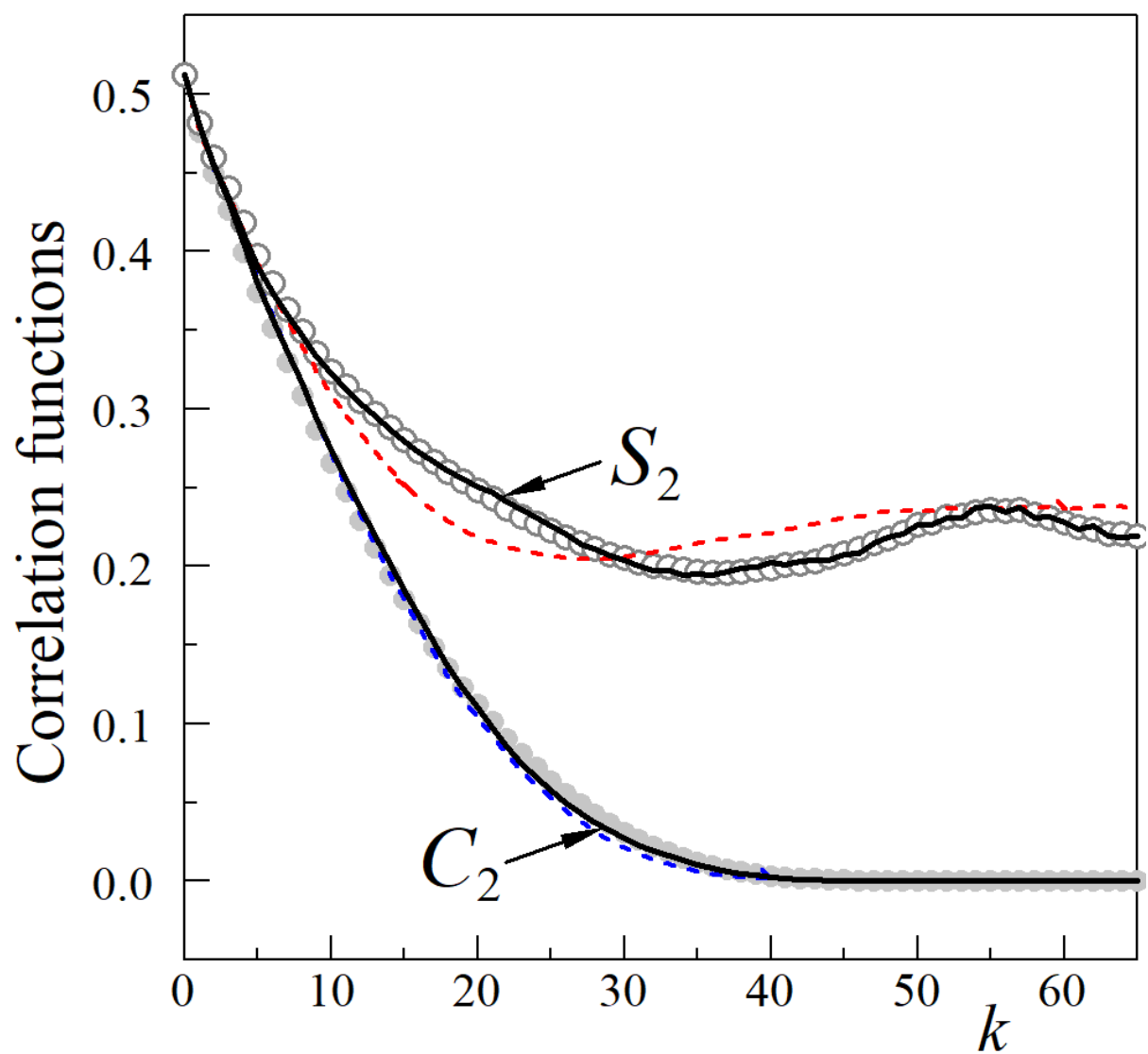

**Fig. 1c**

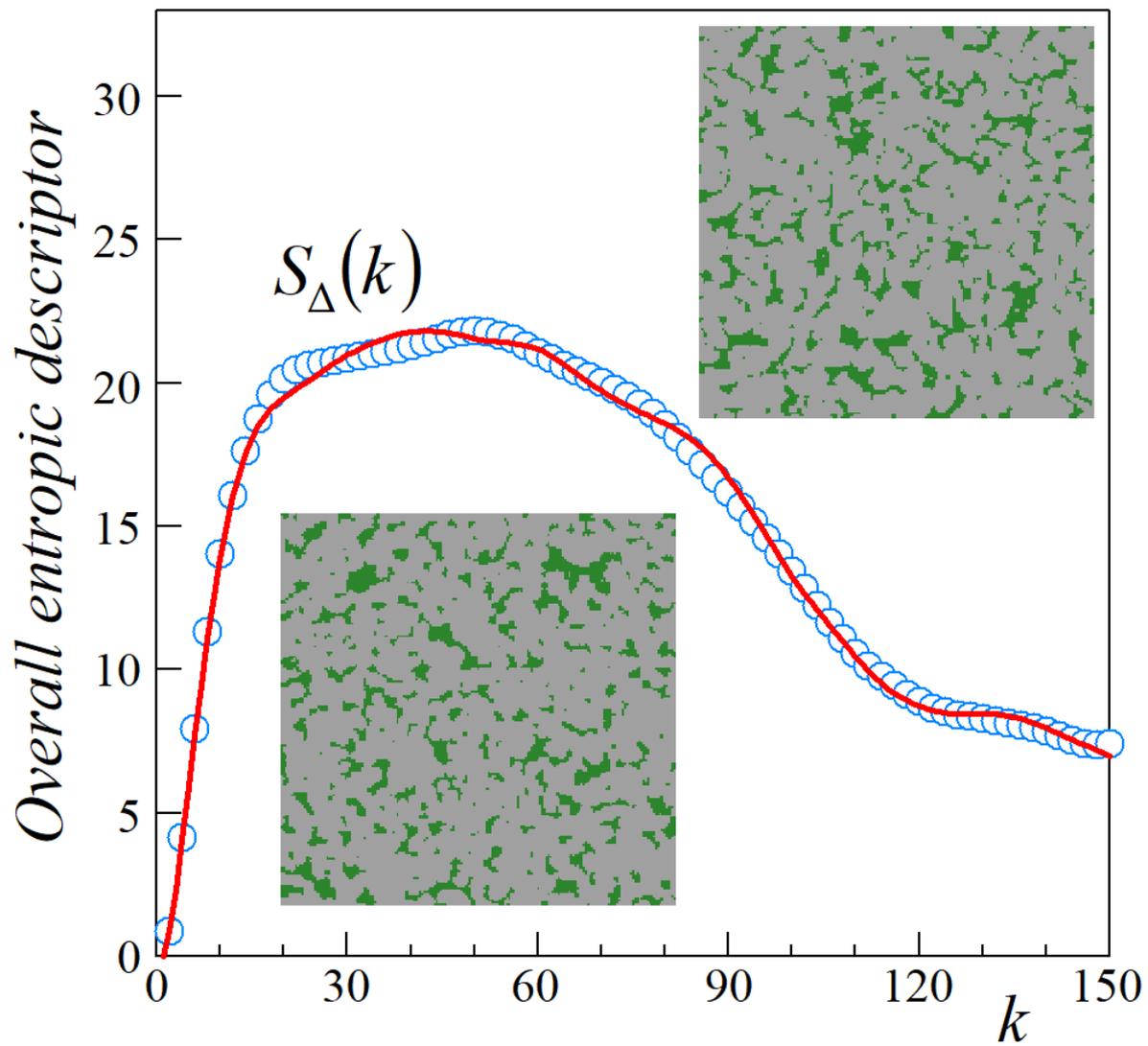

**Fig. 2a**

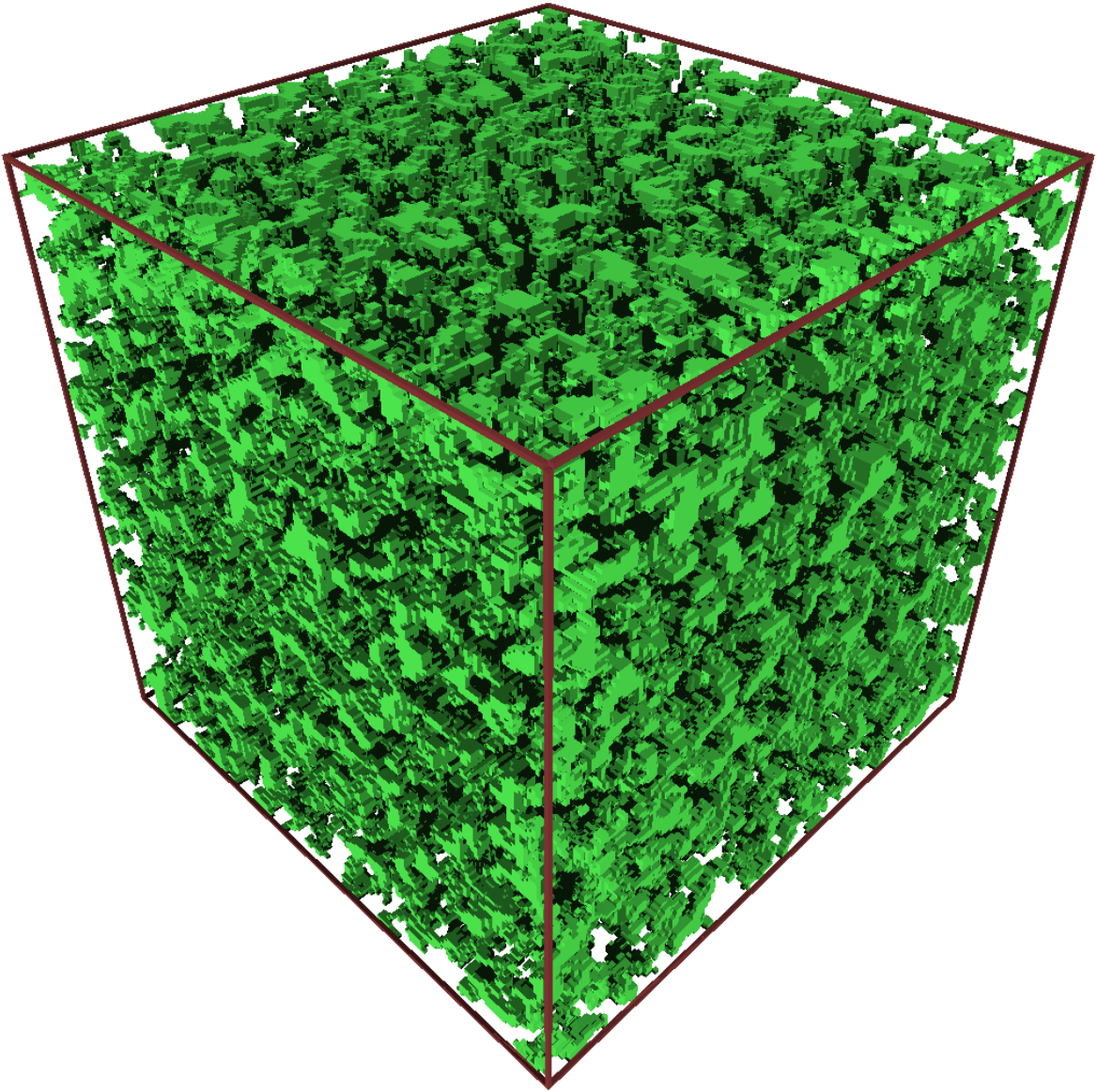

**Fig. 2b**

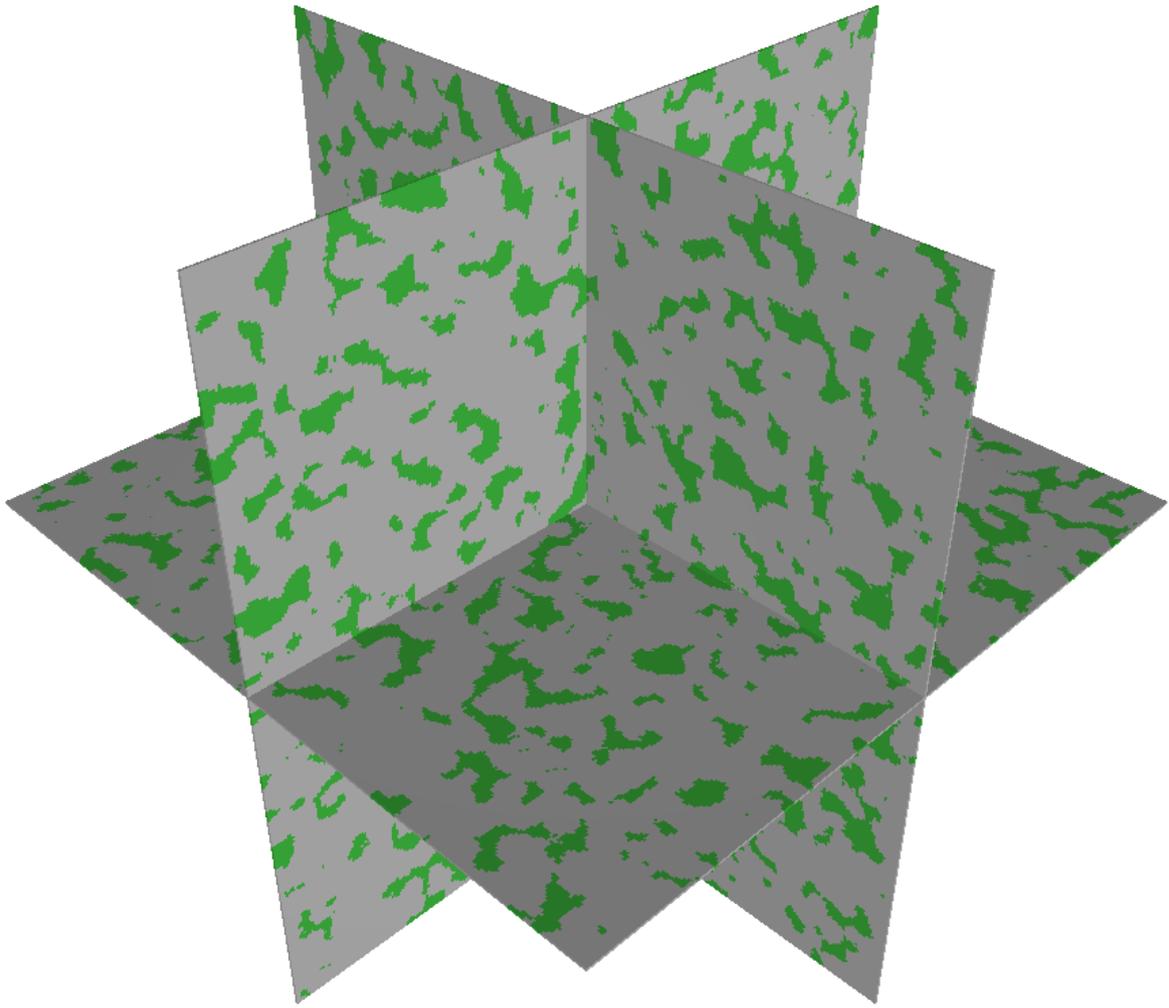

**Fig. 2c**

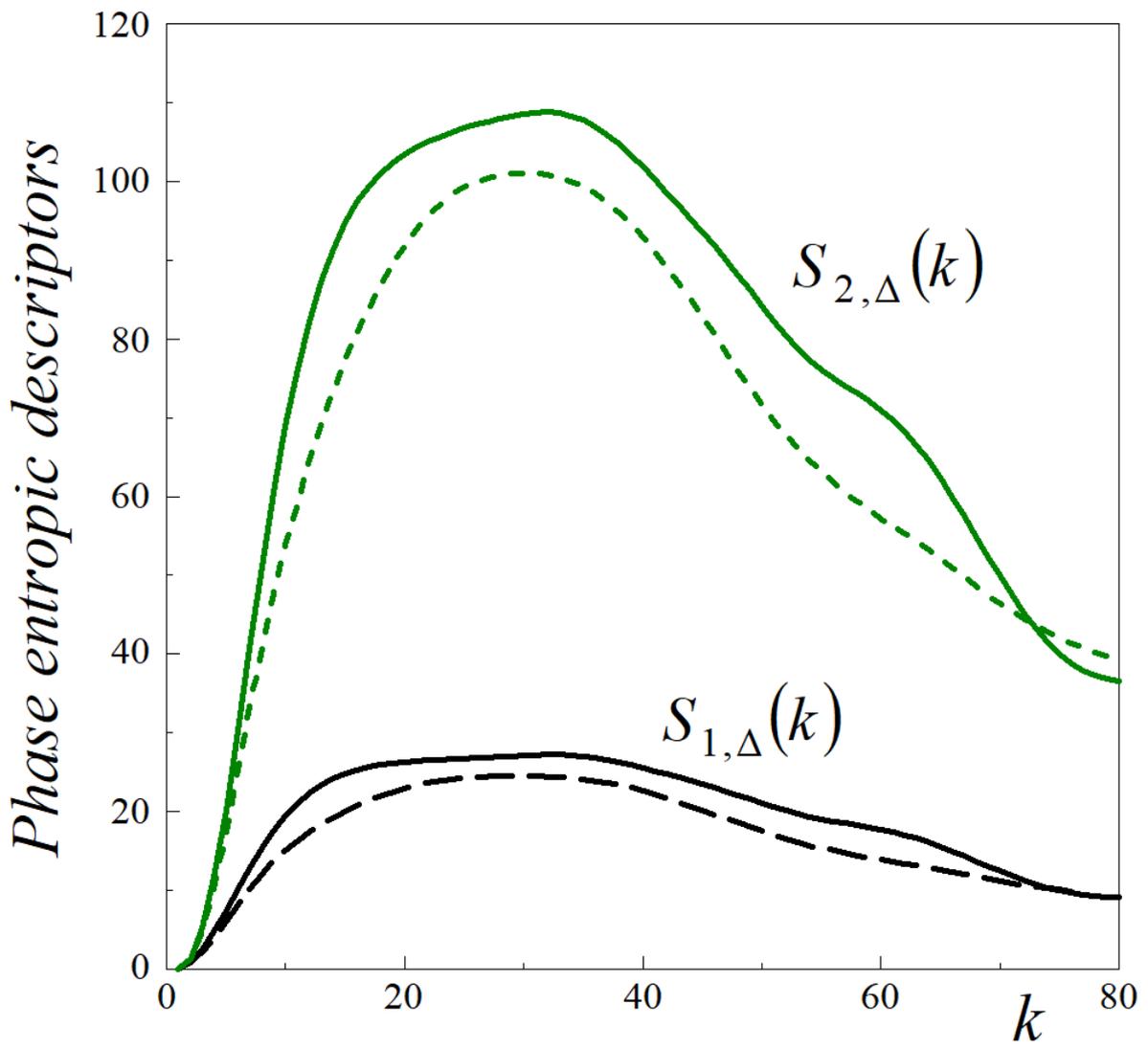

**Fig. 3a**

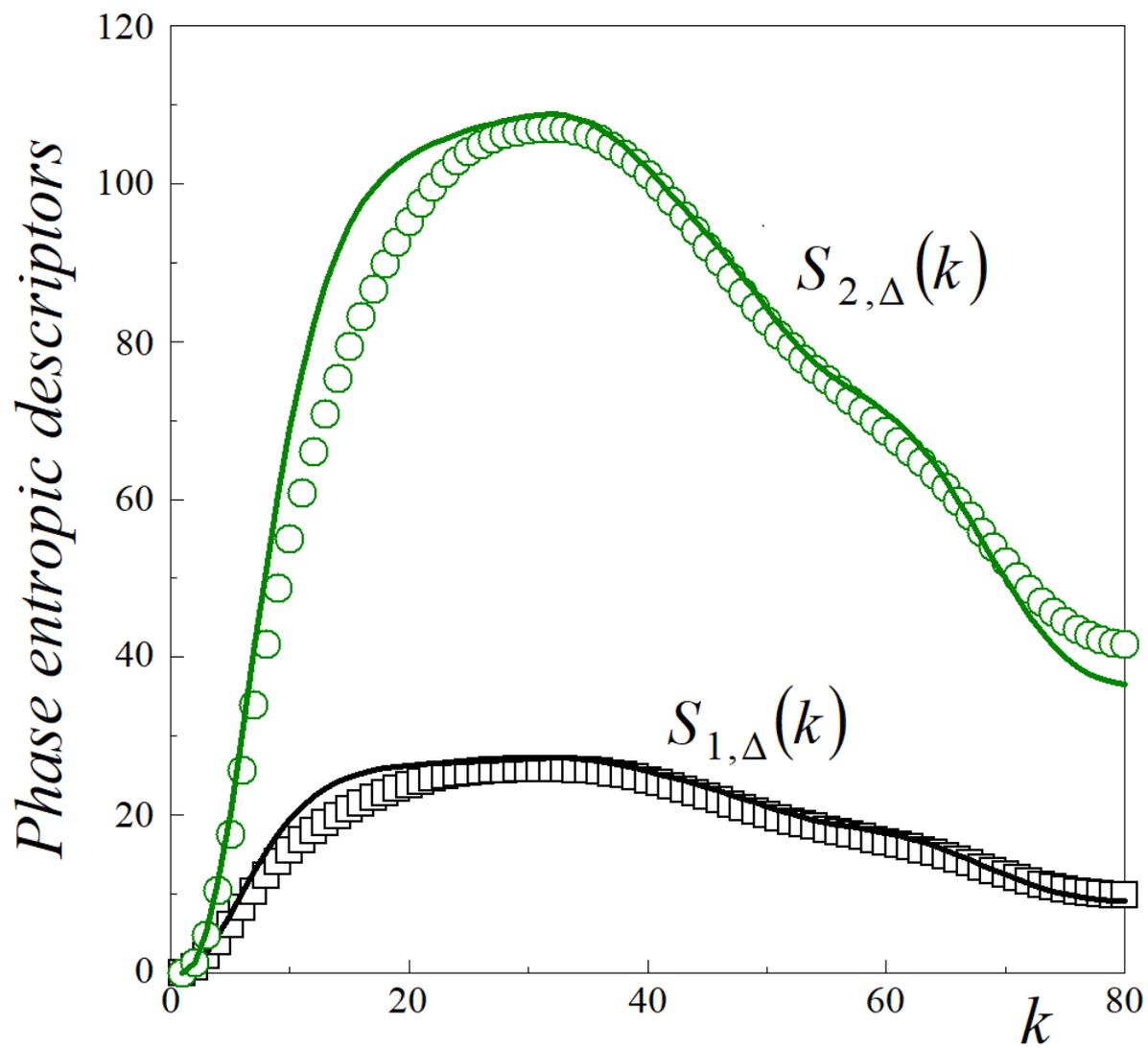

**Fig. 3b**

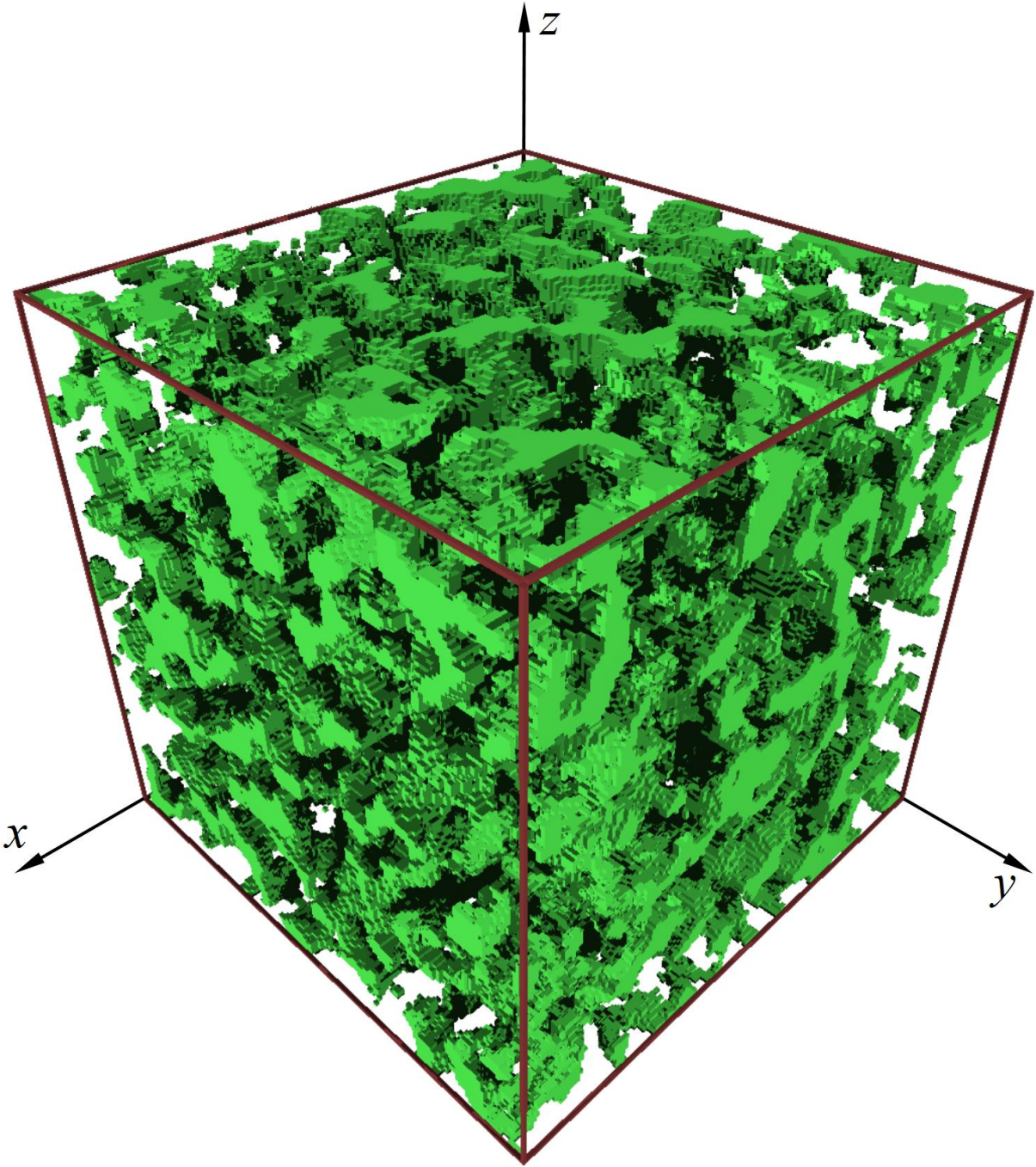

**Fig. 4a**

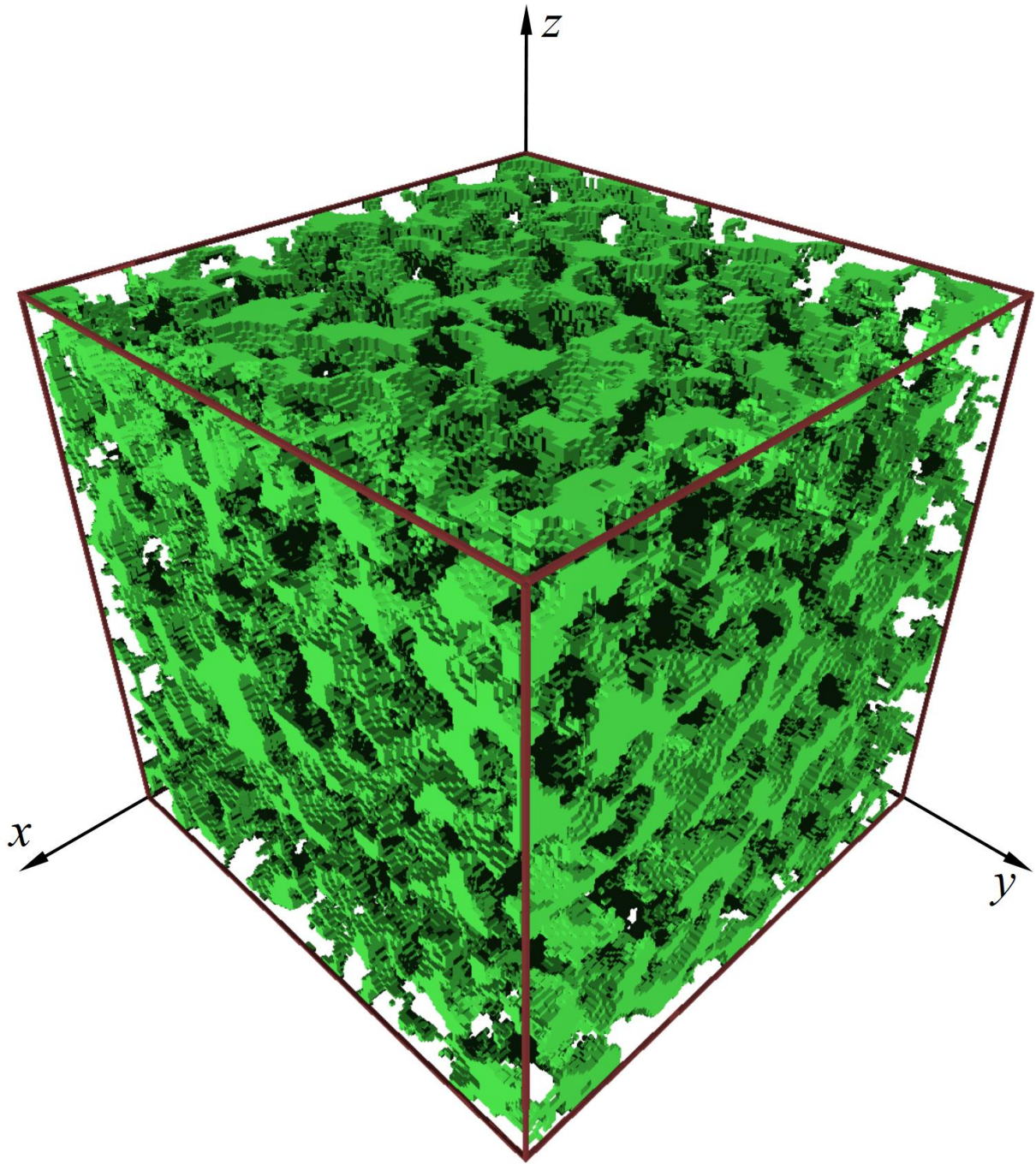

**Fig. 4b**

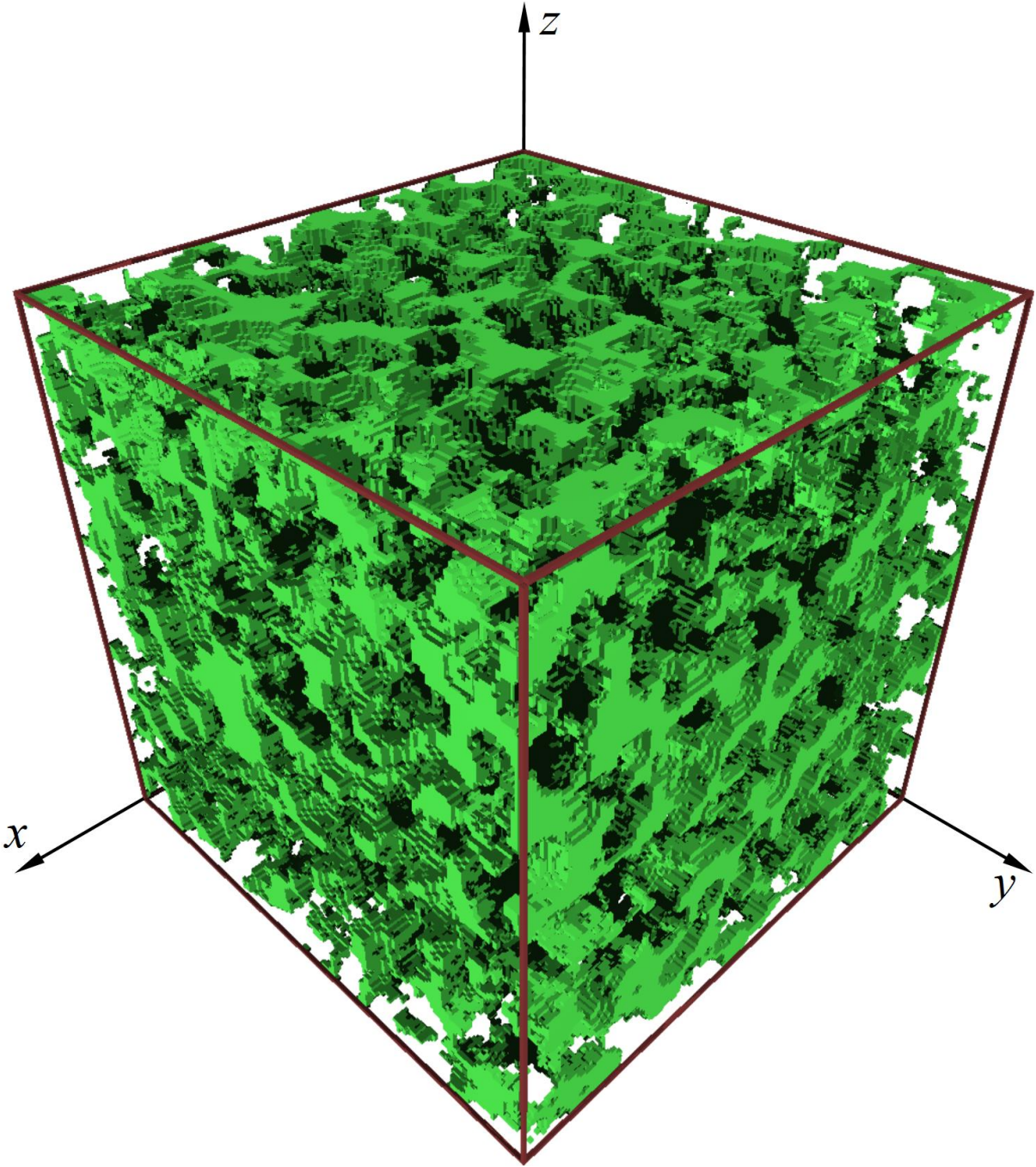

**Fig. 4c**

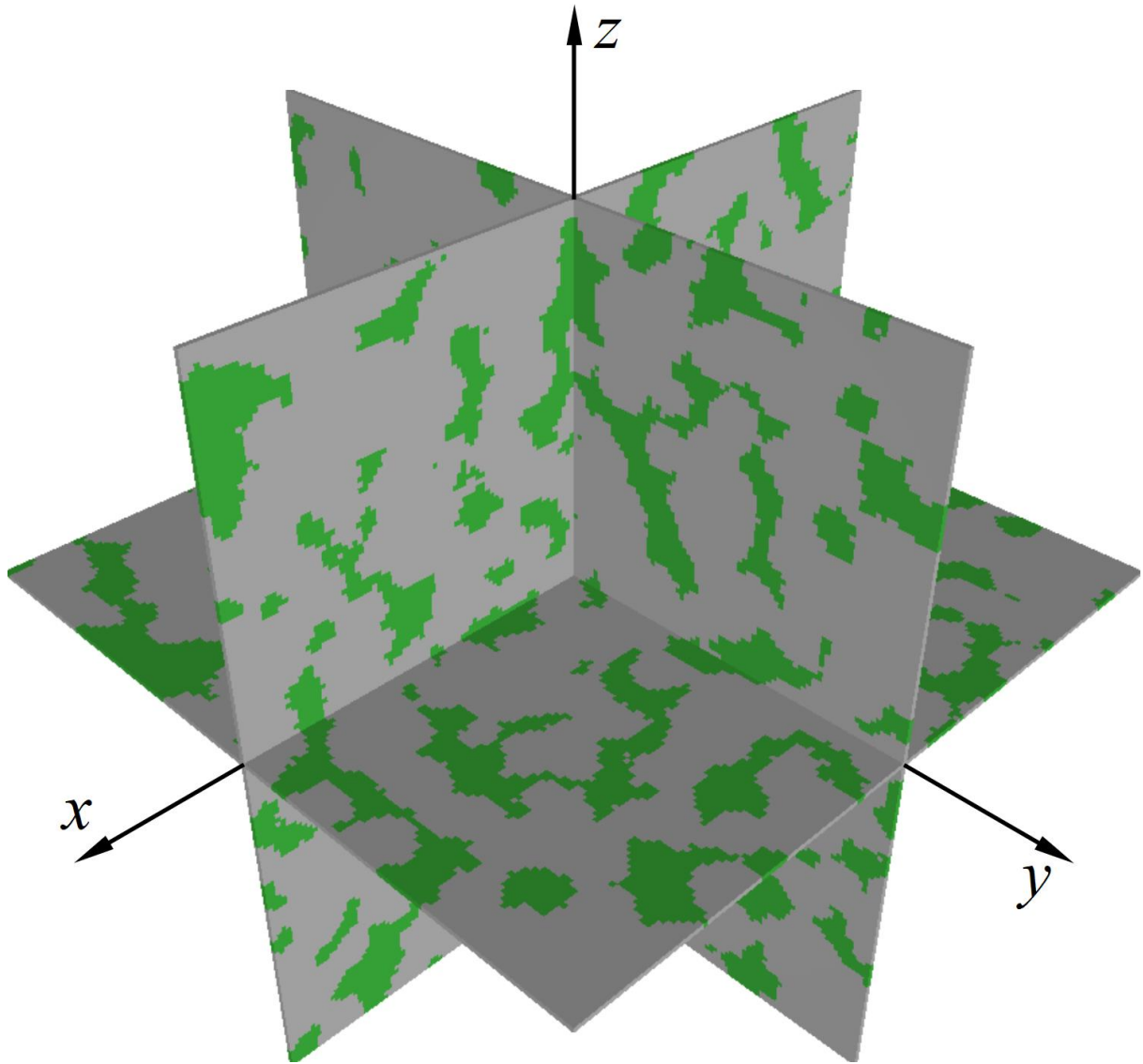

**Fig. 5a**

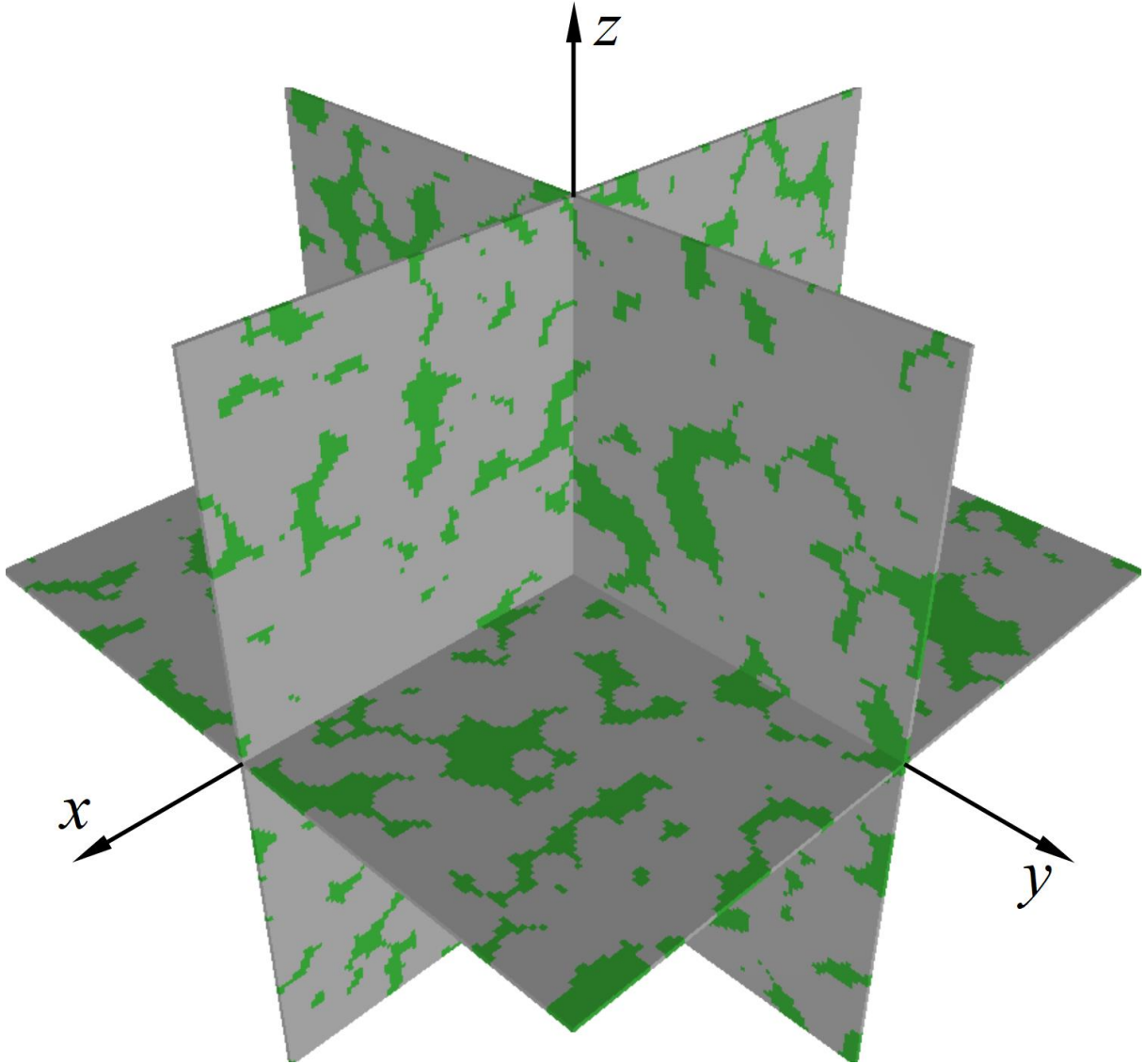

**Fig. 5b**

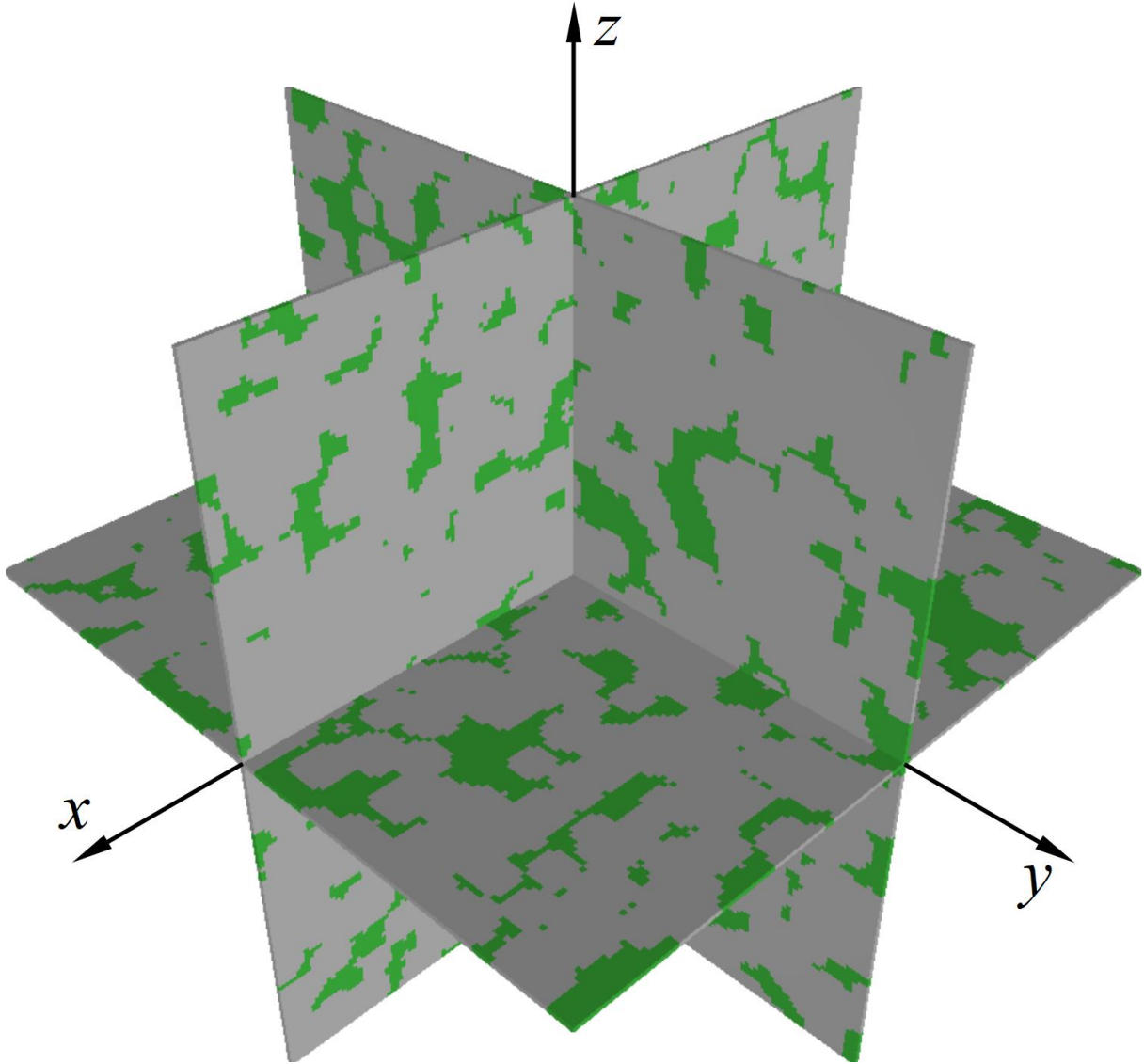

**Fig. 5c**

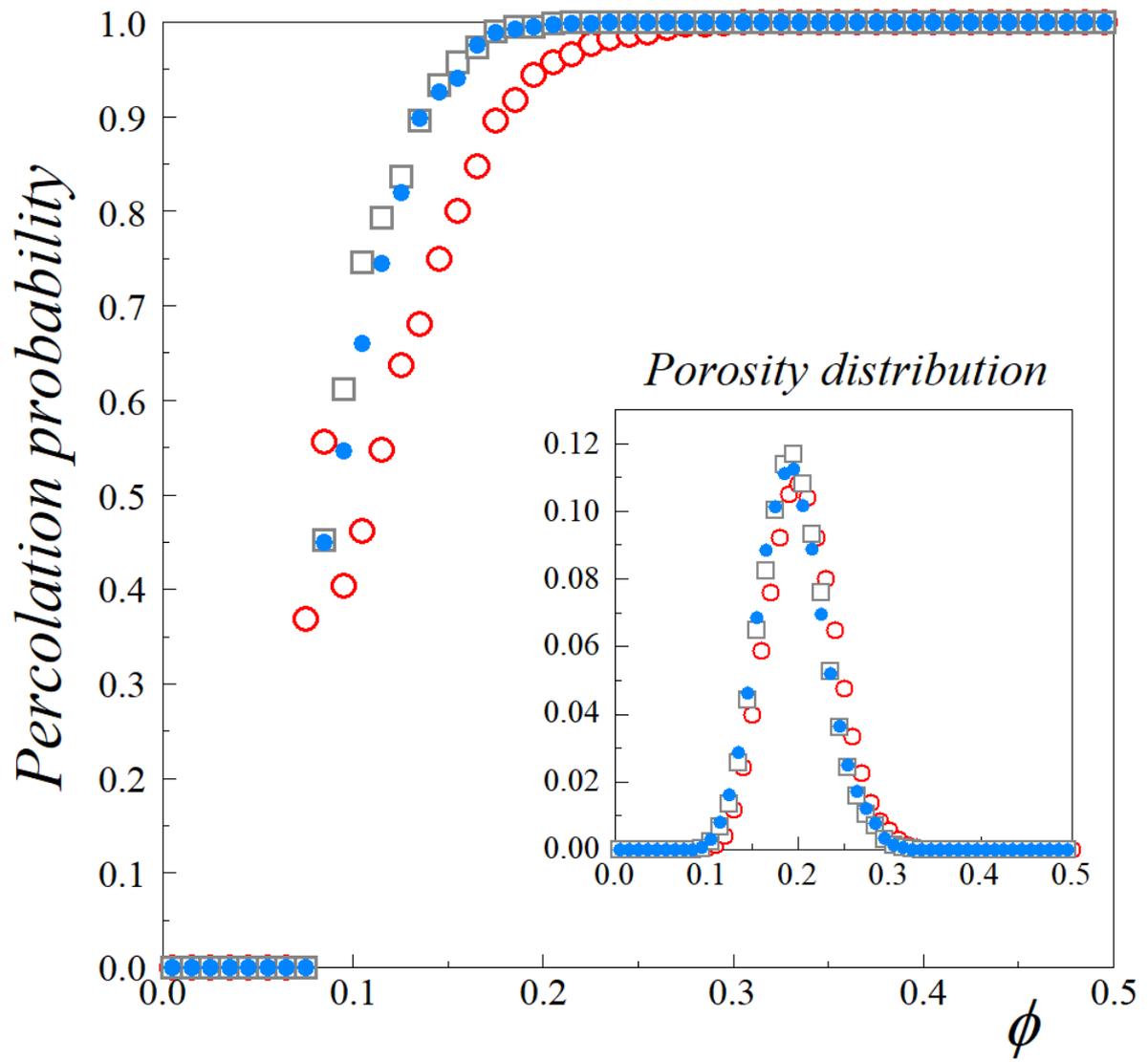

**Fig. 6a**

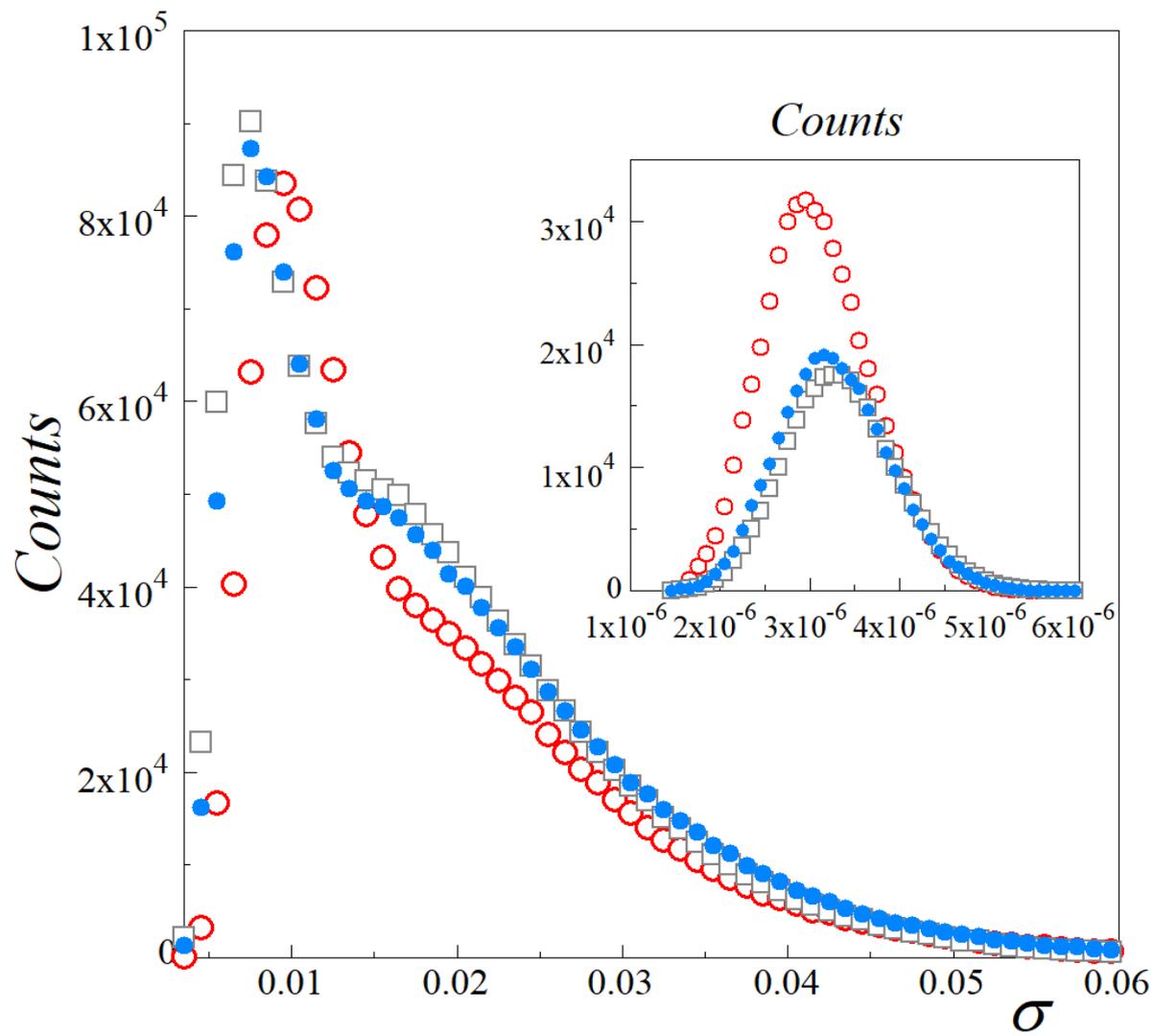

**Fig. 6b**

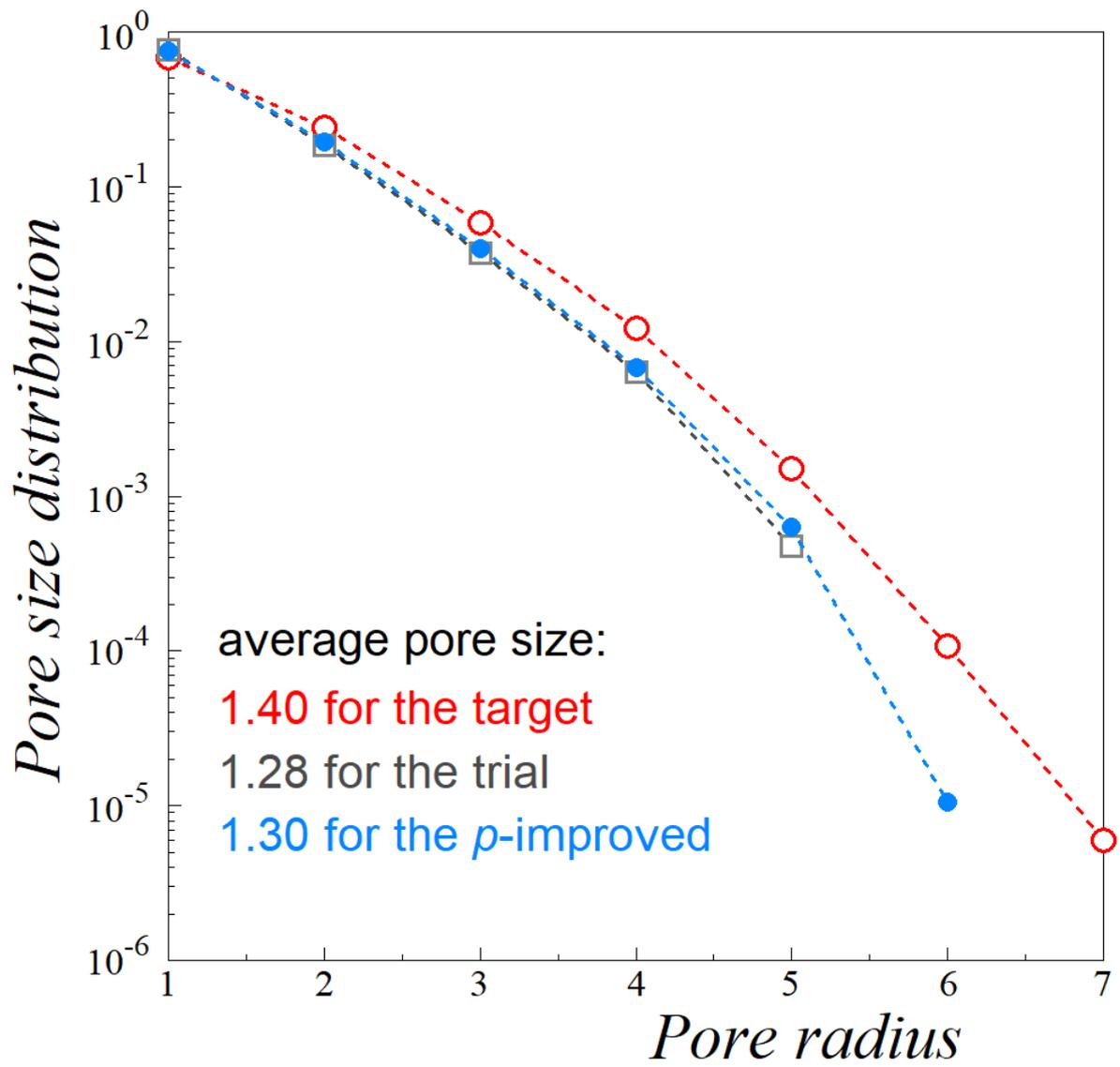

**Fig. 6c**